\documentclass[extra,mreferee] {gji}
\usepackage{times}
\renewcommand{\cite}{\citet}
\usepackage{fancyhdr}
\usepackage{graphicx}
\usepackage{amsmath}
\usepackage{graphicx, subfigure,dsfont} 
\usepackage[latin1]{inputenc}
\usepackage{color}
\begin{document}

\title{Stochastic modeling of regional archeomagnetic series}
\author[G. Hellio, N. Gillet, C. Bouligand, D. Jault]
  {G. Hellio$^1$\thanks{} , N. Gillet$^1$, C. Bouligand$^1$, D. Jault$^1$\\
  $^1$ Univ. Grenoble Alpes, CNRS, ISTerre, F-38 041 Grenoble, France (gabrielle.hellio@ujf-grenoble.fr)  }

\maketitle
\thispagestyle{fancy}

\begin{abstract}
We report a new method to infer continuous time series of the declination, inclination and intensity of the magnetic field from archeomagnetic data. Adopting a Bayesian perspective, we need to specify a priori knowledge about the time evolution of the magnetic field. It consists in a time correlation function that we choose to be compatible with present knowledge about the geomagnetic time spectra. The results are presented as distributions of possible values for the declination, inclination or intensity. We find that the methodology can be adapted to account for the age uncertainties of archeological artefacts and we use Markov Chain Monte Carlo to explore the possible dates of observations. We apply the method to intensity datasets from Mari, Syria and to intensity and directional datasets from Paris, France. Our reconstructions display more rapid variations than previous studies and we find that the possible values of geomagnetic field elements are not necessarily normally distributed. Another output of the model is better age estimates of archeological artefacts.
 \end{abstract}
 

\newpage
\section{Introduction}
From 1840 onward, continuous records from ground-based observatories are available and make it possible to characterize the time derivative of the main field, or secular variation, as a function of length-scale \citep{holme11}.
The spectral properties  of magnetic series obtained from these instrumental records can be transposed into prior information on core processes in the framework of stochastic processes \citep{gillet13}. 

Before the first direct measurements, direction and intensity of the magnetic field can be inferred from remanent magnetization of sediments, volcanic deposits or archeological artifacts. The sparse repartition of archeomagnetic data in space and time and their associated large measurement and dating uncertainties limit our ability to recover the spatio-temporal variations of the geomagnetic field over the past few millennia. 
Strong efforts have nevertheless been made to calculate time-dependent global models of the archeomagnetic field from these data \citep{korte03,korte09,licht13}. 
To take advantage of the large amount of data and the relatively dense temporal coverage available in some areas, for instance in Western Europe \citep{donadini09}, archeomagnetic data are also used to construct regional curves (so-called master curves) that describe the temporal behavior of the magnetic field \citep{legoff02,lanos05,thebault10}. 
Beyond information on processes occurring in the core, master curves provide useful tools for archeomagnetic dating. 

In this study, we focus on the construction of regional archeomagnetic models describing the time evolution of the declination, inclination and intensity of the Earth's magnetic field over the past 6000 years. 
To compensate for the uneven repartition of data and their large uncertainties and to reduce the non-uniqueness, the construction of such models usually incorporates a regularization in time that consists in penalizing the second time derivative of the field \citep{bloxham92}. 
Such regularizations, however, arbitrarily smooth the reconstructed time fluctuations. 
Instead, we rely here on a Gaussian process regression method based on prior information extrapolated from the statistical properties of models obtained from satellite and observatory data. 

Furthermore, dating uncertainties in archeomagnetic data are an important source of errors in the construction of master curves and most inversion methods do not directly account for them. 
Indeed, dating errors are often converted into equivalent measurement errors \citep{korte05}; alternatively, they are estimated using bootstrap or jack-knife methods, which consist in investigating the variability of models obtained from an ensemble of randomly noised and/or sub-sampled datasets \citep{korte09}. 
Here, we use Markov Chain Monte Carlo (MCMC) methods for the dates at which observations have been obtained, based on the probability inherent to the Gaussian process method. 

This paper is divided into 5 sections.
We present in the next section the Gaussian process regression framework, our choice of prior information for the model parameters, the use of Markov Chain Monte Carlo on observation dates and a robust measure of data errors in order to decrease the effect of outliers. 
The method is tested using synthetic observations (section 3), before being applied to datasets from France and the Middle East (section 4). 
A discussion of our results and conclusions are presented in section 5. 

\newpage
\section{Method}
We consider geomagnetic series as the realization of a stochastic process sampled through observations. We use the Gaussian Process Regression (section 2.1) to couple the information contained in measurements with that from the a priori time covariance function of the process. To account for dating errors, we integrate the regression method into a Markov Chain Monte Carlo algorithm, as described in section 2.2. We define, in section 2.3, the a priori information on which relies the Gaussian process framework.  Finally, in section 2.4, we show how to incorporate a robust measure of data errors in order to decrease the effect of outliers that appears when using a standard L2-measure with geophysical series. 

\subsection{Gaussian process regression}

Let us consider a Gaussian, stationary stochastic process $\varphi(t)=\overline{\varphi}+\varphi'(t)$, defined by its average value $\overline{\varphi}$, the perturbation $\varphi'(t)$ from this mean value and its covariance function: 
%\begin{linenomath*}
\begin{equation}
\displaystyle \mbox{Cov}(\varphi(t),\varphi(t+\tau)) = E(\varphi'(t) \varphi'(t+\tau))=  \sigma^2\rho(\tau)\,,
\label{covar GP}
\end{equation}
%\end{linenomath*}
with $\sigma^2$ the variance and $\rho$ the autocorrelation function of the process that will contain the a priori information on the model parameters ; the notation $E(\dots)$ stands for the statistical expectation. 
The continuous process $\varphi$ is sampled with data stored at discrete times into a vector $\bf{y}$, and estimated as a sequence of parameters stored into a vector $\bf{m}=\overline{\bf m}+{\bf m}'$, with $\overline{\bf m}$ the background model and ${\bf m}'$ the model perturbation. 
In our context we consider that the parameters in ${\bf m}$ are homogeneous to the observations in ${\bf y}$ (they are images of the same quantity).
Vectors ${\bf t}_y$ and ${\bf t}_m$ contain respectively the epochs at which the data and the model are sampled. 

The estimate of the model $\bf{m}$, given the data ${\bf y}$ and the measurement errors ${\bf e}$, is characterized by the a posteriori expectation model
%\begin{linenomath*}
\begin{equation}
\displaystyle
\hat{\bf{m}}=\overline{\bf m}+{\sf C}_{my} ({\sf C}_{yy}+{\sf C}_{ee})^{-1}({\bf y - \overline{y}})\,,
\label{expect}
\end{equation}
%\end{linenomath*}
and the a posteriori covariance matrix ${\sf C^*}$: 
%\begin{linenomath*}
\begin{equation}
\displaystyle
{\sf C^*}={\sf C}_{mm}-{\sf C}_{my} ({\sf C}_{yy}+{\sf C}_{ee})^{-1} {\sf C}_{my}^T\,
\label{cpost}
\end{equation}
%\end{linenomath*}
\citep{rasmussen06}. Here $\overline{\bf{y}}$ is the prediction from the background model $\overline{\bf{m}}$ at times ${\bf t}_y$, ${\sf C}_{ee}=E({\bf e}{\bf e}^T)$ is the data error covariance matrix. Matrices ${\sf C}_{mm}$, ${\sf C}_{my}$ and ${\sf C}_{yy}$ are derived from the autocorrelation function using expression (\ref{covar GP}): 
%\begin{linenomath*}
\begin{equation}
\displaystyle
{{\sf C}_{mm}}_{ij}=\sigma^2  \rho({t_m}_i-{t_m}_j) ; {{\sf C}_{my}}_{ij}=\sigma^2  \rho({t_m}_i-{t_y}_j) ; {{\sf C}_{yy}}_{ij}=\sigma^2  \rho({t_y}_i-{t_y}_j) 
\label{Cmd}
\end{equation}
%\end{linenomath*}
Note that the above estimate (\ref{expect}) in term of Gaussian process comes down to calculating the BLUE (Best Linear Unbiased Estimator).

\subsection{Accounting for dating uncertainties with Markov Chain Monte Carlo}
Expression (\ref{Cmd}) assumes that each datum $y_i$ is representative of an epoch ${t_y}_i$. Because dating uncertainties are prominent in archeomagnetic databases, we should consider the probability density function (\textit{pdf}) for the date of the datum $y_i$. 
This distribution depends on the dating method: 
it is generally considered Gaussian for $^{14}$C dating \citep[e. g.][]{aguilar2013}, uniform when the date is estimated from historical or archeological constraints \citep[e.g.][]{genevey03}, or more complex in the case of calibrated $^{14}$C dates \citep{reimer09}. 

\begin{figure}
\begin{center}
\subfigure[]{\includegraphics[width=0.5\linewidth]{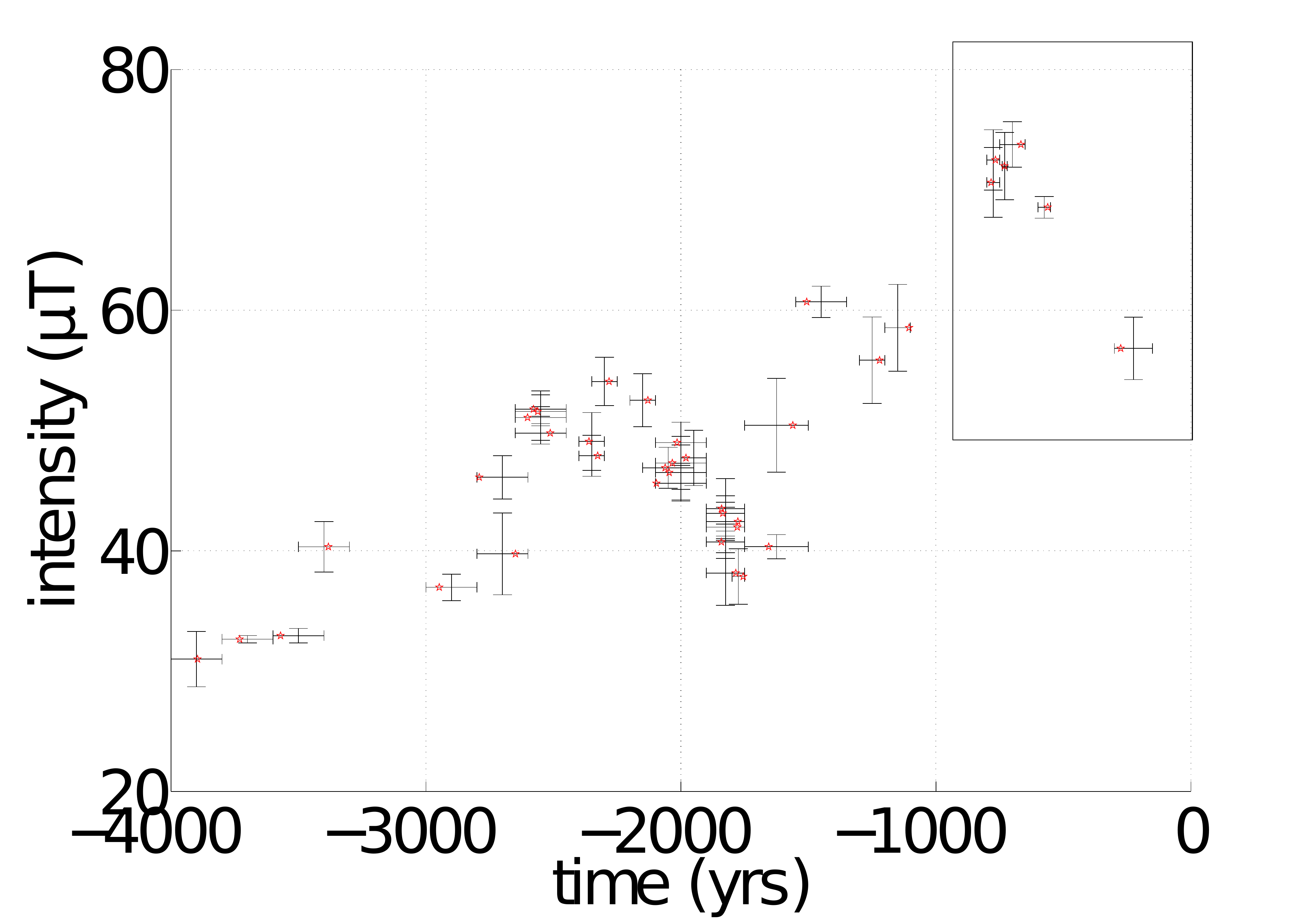}}\\
\subfigure[]{\includegraphics[width=0.5\linewidth]{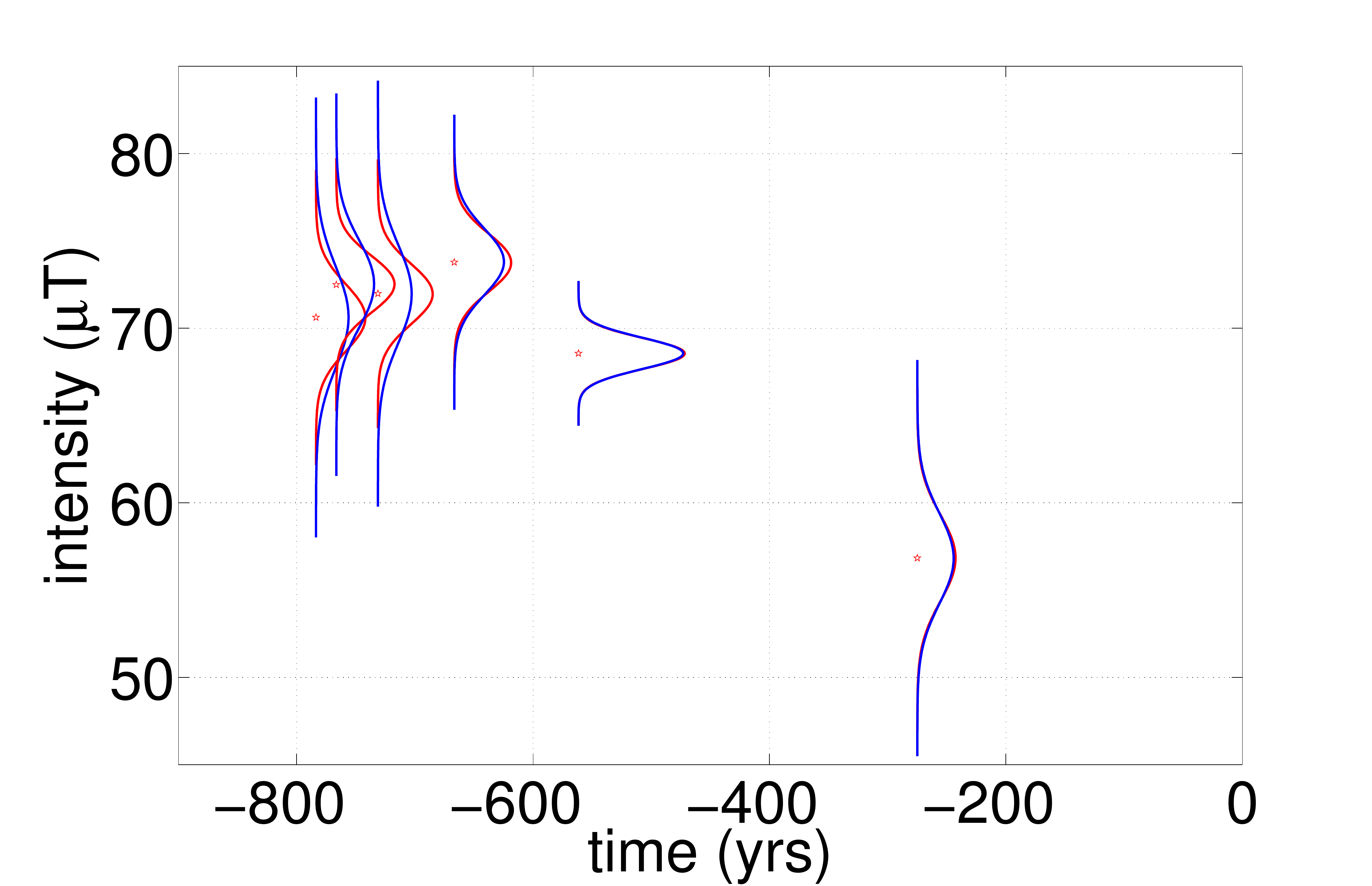}} \\
\subfigure[ ]{\includegraphics[width=0.5 \linewidth]{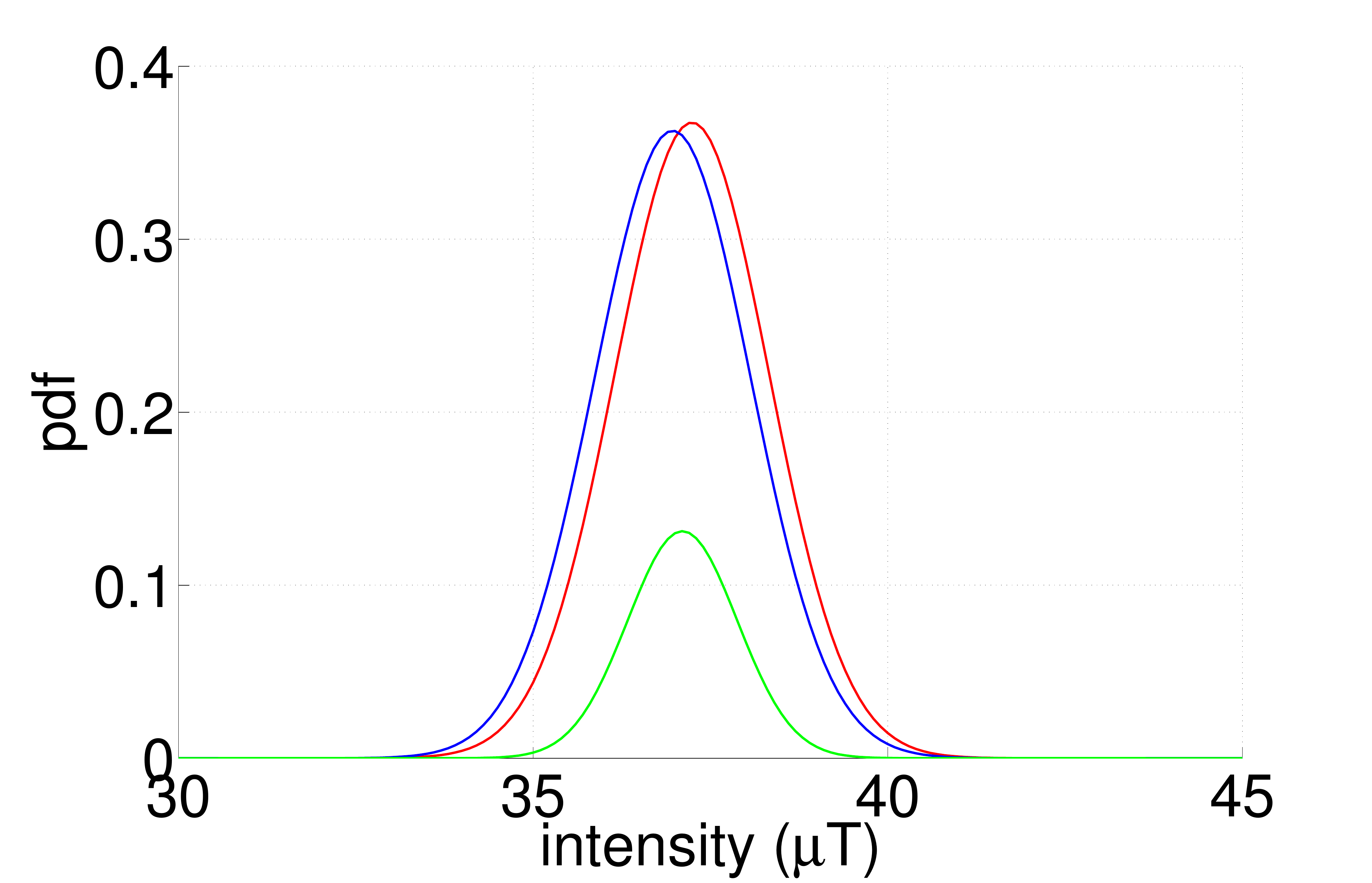}} 
\caption{Illustration of equations (\ref{proba}) and (\ref{proba2}): \label{pdf} a) Syrian intensity dataset (black) together with one draw of random dates (red stars) ; b) Gaussian prior \textit{pdf} of the measurements (blue curves) and Gaussian posterior \textit{pdf} of the model (red curves) for five data from the inset in a) ; c) Gaussian prior \textit{pdf} of the observation at 2900 BC (blue curve), Gaussian posterior \textit{pdf} of the model (red curve), and combined \textit{pdf} of the two (green curve).}
\end{center}
\end{figure}

To consider these dating uncertainties, we build several sets of dates ${\bf t}_y$, illustrated in Figure \ref{pdf}a. We associate at each record a date drawn inside its dating error bar. We estimate for each draw a model ${\bf m}_y$ defined by a mean model ${\bf \hat{m}}_y$ and its covariances ${\sf C}_y^*$ (equations (\ref{expect}) and (\ref{cpost})) at times ${\bf t}_y$. We then evaluate the joint probability of the draw after \cite{lanos04}, see also \citep{pavon11}:

%\begin{linenomath*}
\begin{equation}
p({\bf t}_y,{\bf y}\vert{\bf m}_y) \propto p({\bf m}_y\vert{\bf t}_y,{\bf y}) \times p({\bf t}_y,{\bf y})
\label{proba}
\end{equation}
%\end{linenomath*}

The integration of the probability density function over all possible values of ${\bf y}$ gives the posterior probabilities of the dates ${\bf t}_y$.
%\begin{linenomath*}
\begin{equation}
p({\bf t}_y\vert {\bf m}_y)=\int_{-\infty}^{+\infty} p({\bf t}_y,{\bf y}\vert {\bf m}_y) d {\bf y} \label{proba2}
\end{equation}
%\end{linenomath*}
In practice, we first evaluate these probabilities for each record at time $t_{y_i}$. To this end we multiply the Gaussian posterior probability density function $\mathcal{N}(\hat{m}_{y_i},\sigma_{y_i})$ of the model at time $t_{y_i}$ (red curves in Figure \ref{pdf}b and c), by the Gaussian prior probability density function $\mathcal{N}(y_i,e_i)$ of measurement $y_i$ (blue curves in Figure \ref{pdf}b and c). The notation $\mathcal{N}(\mu,\sigma)$ stands for Gaussian distribution with mean $\mu$ and standard deviation $\sigma$. The standard deviation $\sigma_{y_i}$ of the model at time $t_{y_i}$ is obtained from the posterior covariance matrix ${\sf C}_y^*$.
By this multiplication, we obtain the joint probability density function (green curve in Figure \ref{pdf}c). We then integrate the obtained probability density function over all possible values of $y_i$ to get the probability of the date $t_{y_i}$.
We finally multiply the posterior probabilities of all dates to obtain the probability of draw $k$, noted $P_{\text{draw}_k}$.

A natural way to proceed following \cite{lanos04} is to weigh each mean model given the probability of the corresponding draw. 
However, few draws have very high probabilities compared to all others and numerous iterations provide very few representative mean models. 
To overcome this problem, we use Markov Chain Monte Carlo to explore the possible dates of observations and to select draws with the highest probabilities. We remind here the main steps, (see \cite{aster2013parameter,gilks486markov} for more details):

 1- To explore the possible dates, we generate the $\text{k}_{\text{th}}$ draw from the previous one as a random draw inside the proposal distribution $\mathcal{N}({\bf t}_{k-1},\sigma_{\text{MCMC}})\times p({\bf t})$. For uniform probability distribution, it comes down to a random walk restricted to the a priori time interval.
  
 2- We define an acceptance ratio $\alpha=\text{min}(1,s)$, with $s=\frac{P_{\text{draw}_k}}{P_{\text{draw}_{k-1}}}$.  
 
 3- We keep the $\text{k}_{\text{th}}$ draw if $\alpha>u$, $u$ being a random value obtained from a uniform distribution between 0 and 1, and reject it if not.
 
We stop the chain after $N$ iterations depending on the dataset studied and perform several chains to better explore the space of possible dates \citep[p 13]{gilks1996introducing}. The number of chains is determined by the evolution of the posterior distribution of the dates. The number of accepted draws in each chain depends on $\sigma_\text{MCMC}$. We adjust the latter parameter so the number of kept draws is between 20 and 60$\%$ of all draws. All these informations are summarized in Appendix B. 
Each draw $k$ selected by the above Markov rules consists in a set of dates ${\bf t}_y^k$, associated to records ${\bf y}$ and measurement errors ${\bf e}$. 
From equations (\ref{expect}) and (\ref{cpost}), we obtain for each dataset a mean model ${\bf \hat{m}}$ and its associated covariances ${\sf C}^*$ at the times ${\bf t}_m$.

Expectations and a posteriori covariance matrices are used to build ensembles of models consistent with both the observations and the a priori information assumed for model parameters. 
To this end, we use the Choleski decomposition ${\sf U}$ of the a posteriori covariance matrix, ${\sf C}^*={\sf U}^T{\sf U}$, from which we compute an ensemble of model realizations $\bf{m}=\hat{\bf{m}}+{\sf U}^T \tilde{\bf{m}}$, with $\tilde{\bf{m}}$ a random Gaussian vector with zero mean and unit variance. Each draw selected by the Markov chain is used to build an ensemble of realizations. We put together all these ensembles to build our final estimate of the probability density function. Note that this distribution is not necessarily Gaussian.

\subsection{A priori covariances on geomagnetic series} \label{sec: series prior}
\label{sec: prior}
We detail below how we derive our covariances on geomagnetic series (intensity $F$, inclination $I$, declination $D$) from a priori covariances on the Gauss spherical harmonic coefficients. These are chosen to be compatible with the temporal power spectral densities recorded in ground-based observatories \citep{gillet13}.

We assume that all Gauss coefficients $(g_n^m, h_n^m)$, with $n$ and $m$ the spherical harmonic degrees and orders, result from an auto-regressive (AR) process of order 2, with correlation function 
%\begin{linenomath*}
\begin{equation}
\rho_n(\tau)=\left(1+\frac{\sqrt{3}\tau}{\tau_c(n)}\right) \exp\left(-\frac{\sqrt{3}\tau}{\tau_c(n)}\right)\,. \label{corr_func}
\end{equation}
%\end{linenomath*}
Covariances for Gauss coefficients are then:
%\begin{linenomath*}
\begin{equation}
\displaystyle
\mbox{Cov}(g_n^m(t),g_n^m(t+\tau))=\sigma^2_g(n)  \rho_n(\tau)=K_n(\tau)\,,
\label{covar gnm}
\end{equation}
%\end{linenomath*}
with a similar notation for $h_n^m$ coefficients. The time $\tau_c$ and the variance $\sigma_g^2$ are functions of the degree $n$ only.
We assume that there is no cross-correlations between Gauss coefficients of different degrees and orders, and between $g$ and $h$ as well.
Note that this correlation function is solution of the stochastic differential equation \citep{yaglom04}:
%\begin{linenomath*}
\begin{equation}
d\frac{d{\varphi'}}{dt}  +\frac{2 \sqrt{3}}{\tau_c}d\varphi' + \frac{3}{\tau_c^2}\varphi' dt= d\zeta(t) \,,
\label{stoch}
\end{equation}
%\end{linenomath*}
where $\zeta(t)$ is the Brownian motion (or Wiener process). 

Variances $\sigma^2_g(n)$ for the non-dipole Gauss coefficients are obtained from the variance of the Gauss coefficients estimated in satellite field models, as in the models COV-OBS \citep{gillet13}: 
%\begin{linenomath*}
\begin{equation}
\sigma_g^2(n)= \frac{1}{2n+1}\sum_{m=0}^n\left[ g_n^m(t)^2+h_n^m(t)^2\right]\,.
\label{gnm var}
\end{equation}
%\end{linenomath*}
Using a similar definition for $\sigma_{\dot{g}}^2(n)$, equation (\ref{stoch}) imposes the value of the correlation time:
%\begin{linenomath*}
\begin{equation}
\displaystyle
\tau_c(n)=\sqrt{3} \frac{\sigma_g(n)}{\sigma_{\dot{g}}(n)}\,.
\label{tau n}
\end{equation}
%\end{linenomath*}
The background model is composed of the axial dipole value $\overline{g}_1^0=-35\mu$T, and the variance for the dipole coefficients is chosen as $\sigma^2_g(1)=5 \mu$T$^2$, the value typically found for the past 4000 years \citep{korte11}. 
Since $\sigma^2_{\dot{g}}(n)$ is not affected by the presence of a stationary background, we find 
a correlation time of about 200 years  for all coefficients of degree one.

We have propagated this a priori information on Gauss coefficients to geomagnetic series of declination $D$, inclination $I$ and intensity $F$ recorded at the Earth's surface. Our approach requires that these quantities have a Gaussian distribution.
It has been shown that the intensity distribution was close to a Gaussian distribution in the limit of small relative dispersion \citep{love03}. 
This is indeed the case for archeomagnetic data, since on centennial to millennial time-scales the standard deviation in the axial dipole is small compared to the average value. 
Assuming that Gauss coefficients are the result of a random stationary process and that they have a zero mean except for the axial dipole $g_1^0$, we show in Appendix A how to obtain the mean, covariance and cross-covariance of geomagnetic series of $D$, $I$ and $F$ (equations (\ref{covar DIF app}) and (\ref{cross-covar DIF app})). Covariances depend on the colatitude $\theta$ of the sampled site, on $\bar{g_1^0}$ and on sums over degree $n$ of the correlation function defined in equation (\ref{corr_func}). Note in particular that we find non-zero covariances between $F$ and $I$.

Studies carried out on magnetic series from paleomagnetic to archeomagnetic records suggest a continuous spectrum of the Virtual Axial Dipole Moment \citep{constable05,ziegler11}, with slope decreasing from about zero on the longest periods towards about -2 at millennial periods. 
The analysis of models of Holocene lake sediment magnetic records \citep{panovska2013observed}
has shown that temporal power spectra for declination, inclination and relative paleointensity from lake sediments data follow a power law with a slope $-2.3\pm0.6$ for periods between 300 and 4000 years. 
These findings are in good agreement with recent results obtained for the dipole moment from geodynamo numerical simulations \citep{olson12}, which also display steeper slopes at higher periods. 

\begin{figure}
\begin{center}
\includegraphics[width=1.\textwidth]{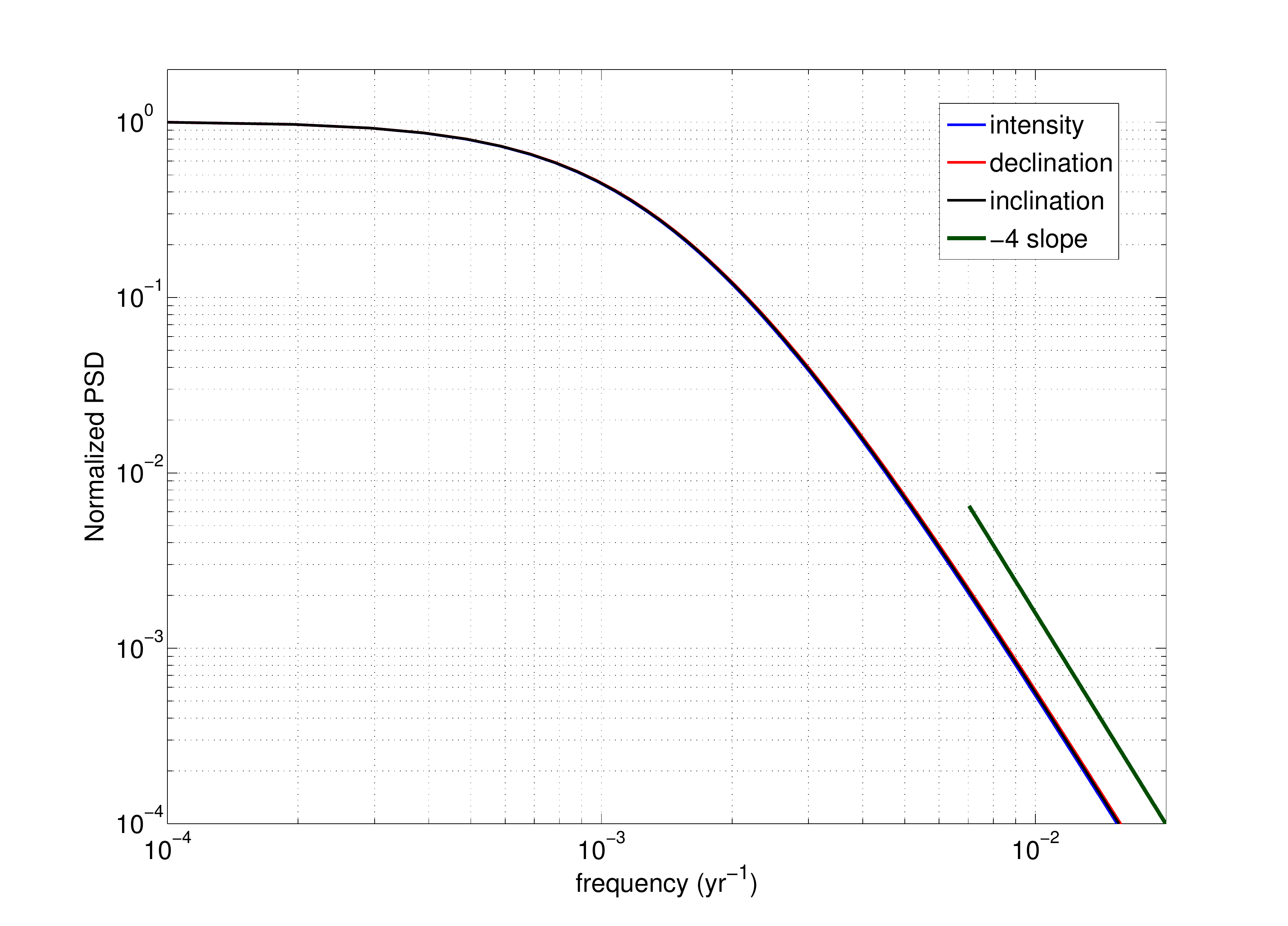}
\caption{Normalized power spectral density of intensity (blue), inclination (black) and declination (red) calculated at co-latitude $\theta=45°$ with a spherical harmonic truncation $N=14$. The -4 power law is plotted for comparison in green. Note that curves are superimposed. \label{norm_spec}}
\end{center}
\end{figure}

The a priori information discussed above presents the advantage to require only a single parameter per degree ($\tau_c$). The slope of the temporal power spectrum for a process defined by equation (\ref{stoch}) is by construction -4 at periods $\tau\ll\tau_c$ \citep{gillet13}, which agrees with that obtained for observatory series \citep{desantis03}. We illustrate in Figure \ref{norm_spec} that we retrieve the -4 slope for spectra of the auto-correlation functions for $F$, $D$ and $I$, obtained with equation (A.15) -- the square of the power spectrum for a series $\varphi(t)$ is the power spectrum of its covariance function ${\sf Cov}(\varphi(t);\varphi(t+\tau))$. The choice of a priori information in the present study is particularly important for periods shorter than a few hundred of years. Indeed, archeomagnetic data being sparse in time, it is towards high frequencies that we need to buttress the evidence from observations with prior information.

\subsection{Dealing with outliers using robust measures of the data errors}
The methodology developed hitherto relies on a L2-norm to account for measurement errors, which makes the approach vulnerable to large errors. Outliers to the L2-norm are unfortunately a common feature of archeomagnetic data analyses \citep{suttie11}. To decrease the effect of these outliers when using L2-norms,  \cite{donadini09} assigned to all data a minimum value for the measurement errors (5$\mu$T for intensity data and 4.3$°$ for directional data). We can instead modify the measure of the misfit to observations and replace the L2-norm with the Huber norm, which distribution is defined as:  (see \cite{farquharson98}):
%\begin{linenomath*}
\begin{equation}
  p(r)=\frac{1}{N} \left \{
   \begin{array}{l c l l r}
         \exp(-\frac{r^2}{2}) & \; , & |r| & < & c \\
     \exp(-c |r| + \frac{c^2}{2})  & \; ,  & |r| & \ge & c
        \end{array}
   \right .,
   \label{huber}
\end{equation}
%\end{linenomath*}
with $N=2.6046$ for $c=1.5$ in this study and $r$, the normalized data misfit residuals. To implement the Huber norm with the previous method, we use the iteratively re-weighting least-squares algorithm where the matrix  ${\sf C}_{ee}$ is constructed  from the residual of the data $i$, $r_i=\dfrac{|y_i - \hat{m}_{yi}|}{\sigma_{yi}}$, as
%\begin{linenomath*}
\begin{equation}
  {\sf C}_{eeii}=  \left \{
   \begin{array}{l c l l r}
     \sigma_{yi}^2 & \; , & r_i & < & c \\
     \dfrac{ r_i \sigma_{yi}^2}{c}  & \;  ,& r_i & \ge & c 
        \end{array}
   \right . .
\end{equation}
%\end{linenomath*}
The Huber norm impacts also the joint probability (equation (\ref{proba})). Instead of multiplying the Gaussian posterior probability density function of the model by the Gaussian prior probability density function of the measurements, we multiply it by the Huber probability density function defined in equation (\ref{huber}). Few iterations are needed to obtain convergency.  The use of the Huber norm rescales the weight in ${\sf C}_{ee}$ associated with outliers. We present in the following section synthetic tests for which there is no need to use this norm since there are no outliers. In section 4 however, we apply the Huber norm to all geophysical datasets. We compare it with the L2-norm for the Syrian series to show how it reduces the effect of outliers. 
\newpage
\section{Synthetic tests}

In order to test the Gaussian process regression on observations presenting dating errors (accounted for with the MCMC method), we build synthetic datasets of $D$, $I$ and $F$ that are consistent with an AR process of order 2 as defined in section \ref{sec: prior}, and that display similar characteristics to real archeomagnetic datasets in terms of temporal distribution and errors. 
To this end, we first construct series for the period 3000 BC to 2000 AD, sampled every 10 years, using the covariance functions defined in equations (\ref{covar DIF app}) and (\ref{cross-covar DIF app}). 
In these covariance functions, the functions $K_n(\tau)$ are defined using the variances and correlation times defined in equations (\ref{gnm var}) and (\ref{tau n}), and the sums are performed with a spherical harmonic truncation degree $N=14$. 
We observe that the model is not modified when increasing further this truncation degree, and that it is already converged with $N=4$. 
We then randomly sub-sample the series and add random measurement and dating errors to each data. 
These errors are built using a Gaussian law for measurement errors and a uniform law for dating errors, to mimic the dating uncertainties from historical constraints. We present for comparison the results considering Gaussian dating errors. 
Finally, the measurement and dating errors used in the modeling phase correspond to the standard deviation and the half-width of the law used to build them. We report in Appendix B, the parameters used for MCMC method for all studied series.
We use two different datasets consisting of 20 and 50 records respectively with randomly assigned dating and measurement errors. 
Dating errors are generated from a uniform distribution with a half-width of 25 years, and measurement errors from a Gaussian distribution with a standard deviation of 1$\mu$T. 

\begin{figure}
\begin{center}
\subfigure[]{\includegraphics[width=0.49\linewidth]{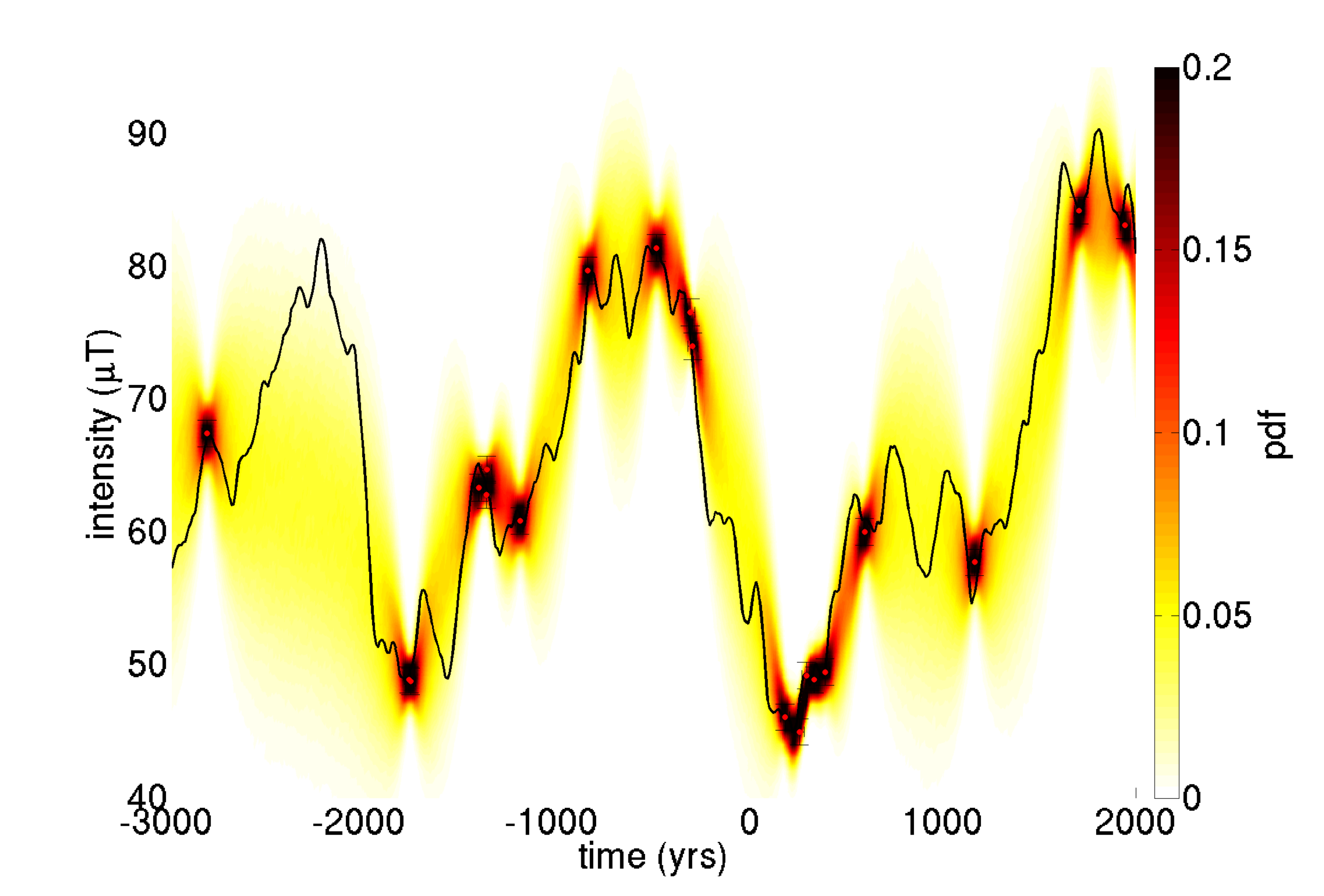}\label{F_10}}
\subfigure[]{\includegraphics[width=0.49\linewidth]{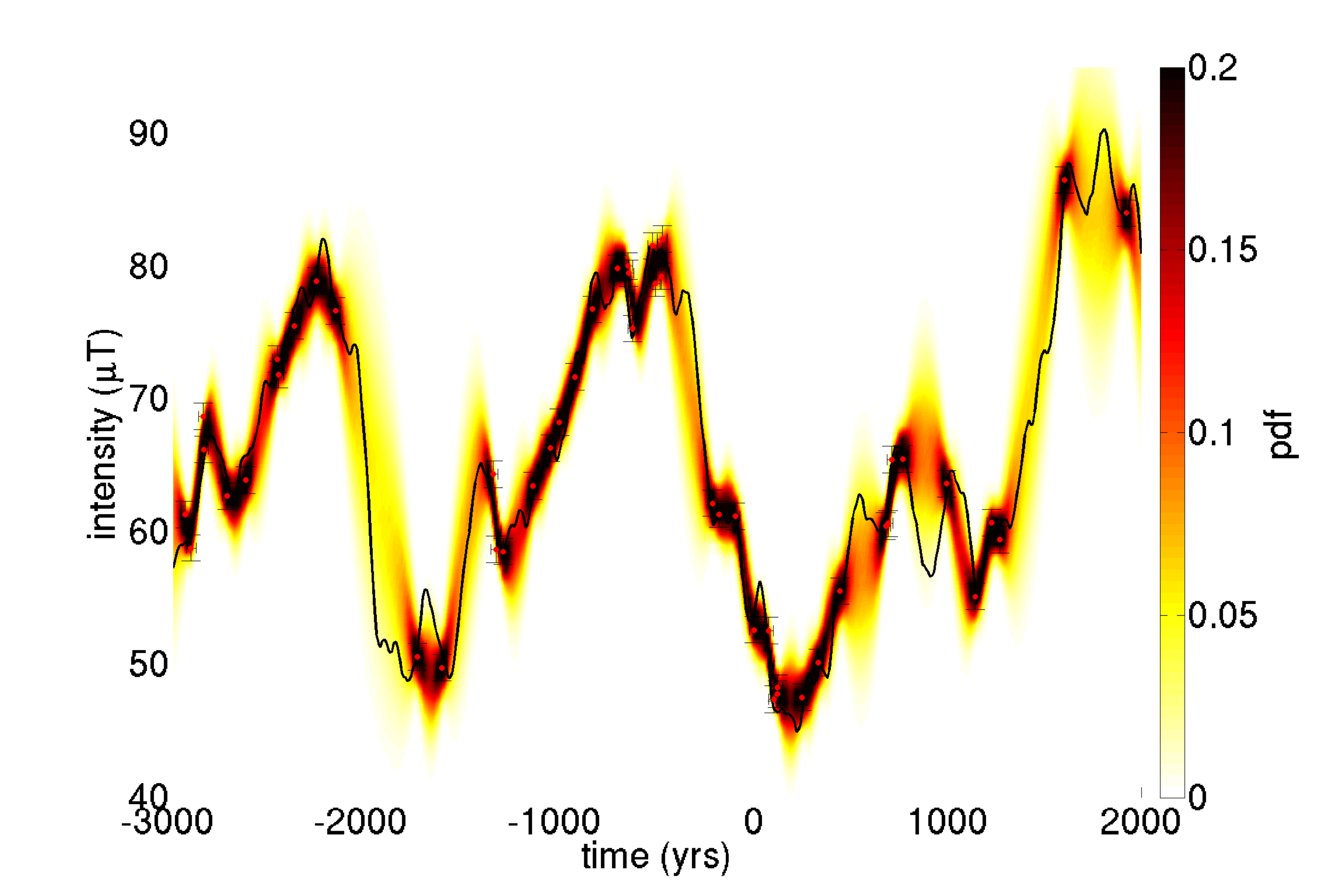}\label{F_50}} \\
\subfigure[]{\includegraphics[width=0.49\linewidth]{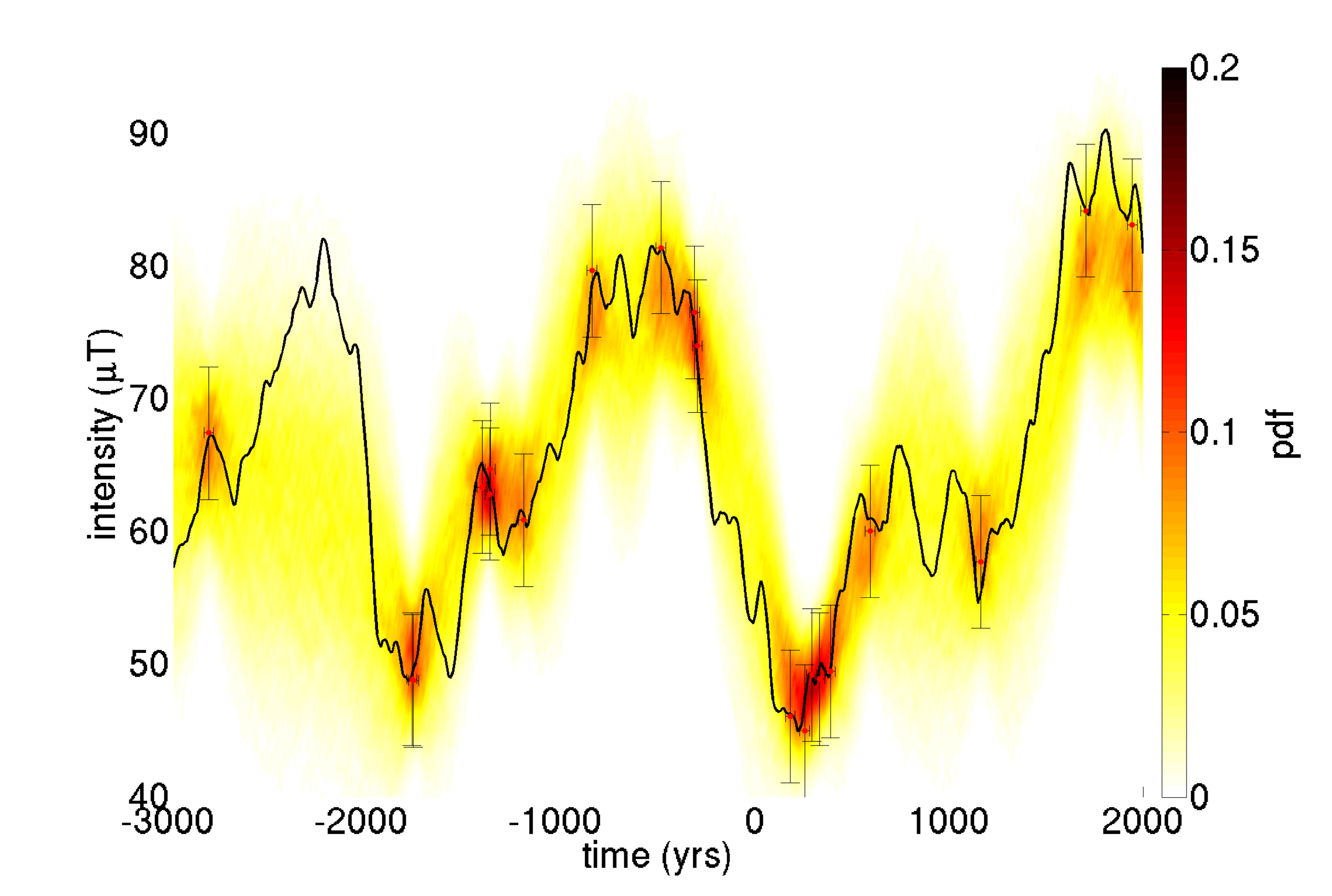}\label{F_more}}
\subfigure[]{\includegraphics[width=0.49\linewidth]{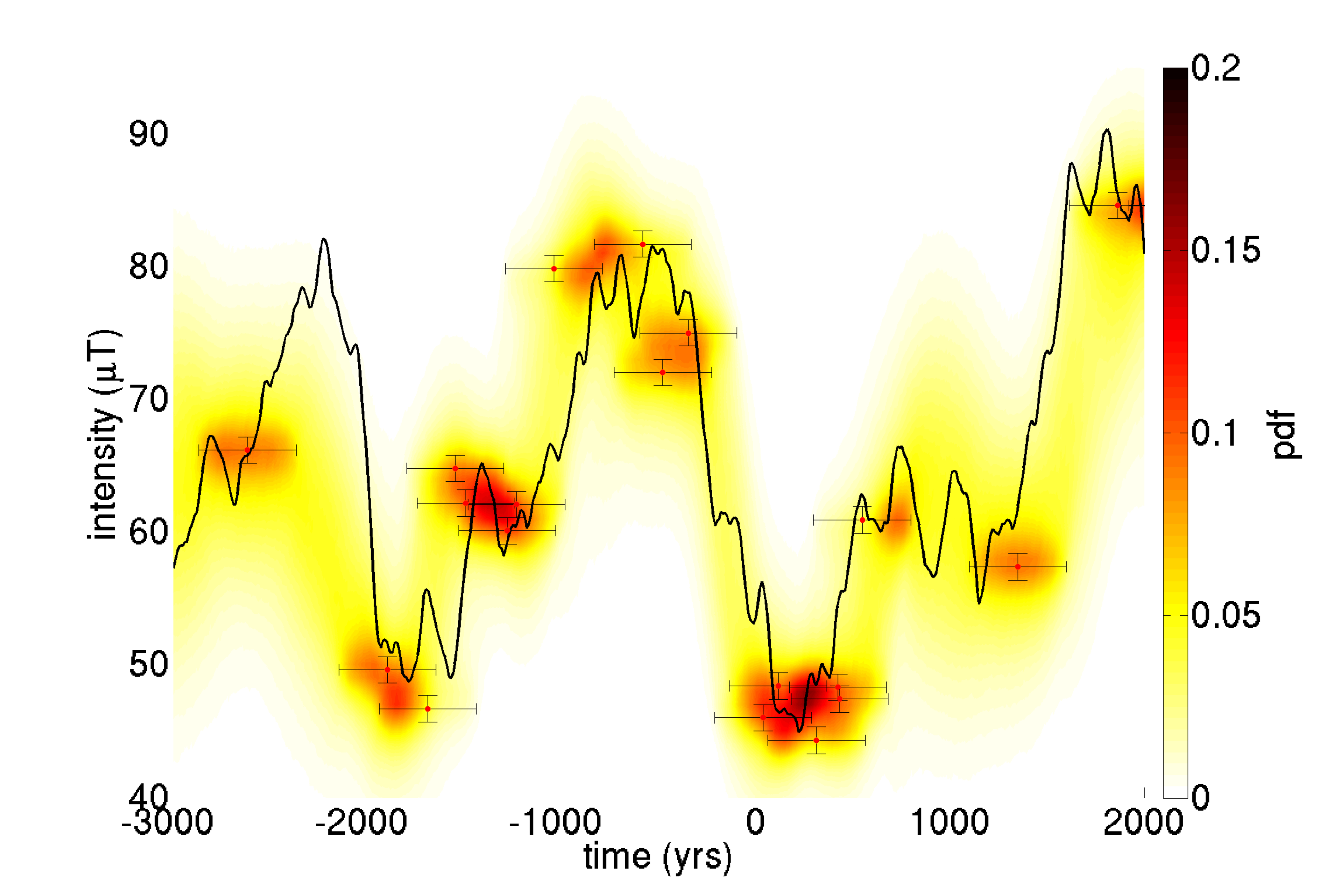}\label{F_10times}}
\subfigure[]{\includegraphics[width=0.49\linewidth]{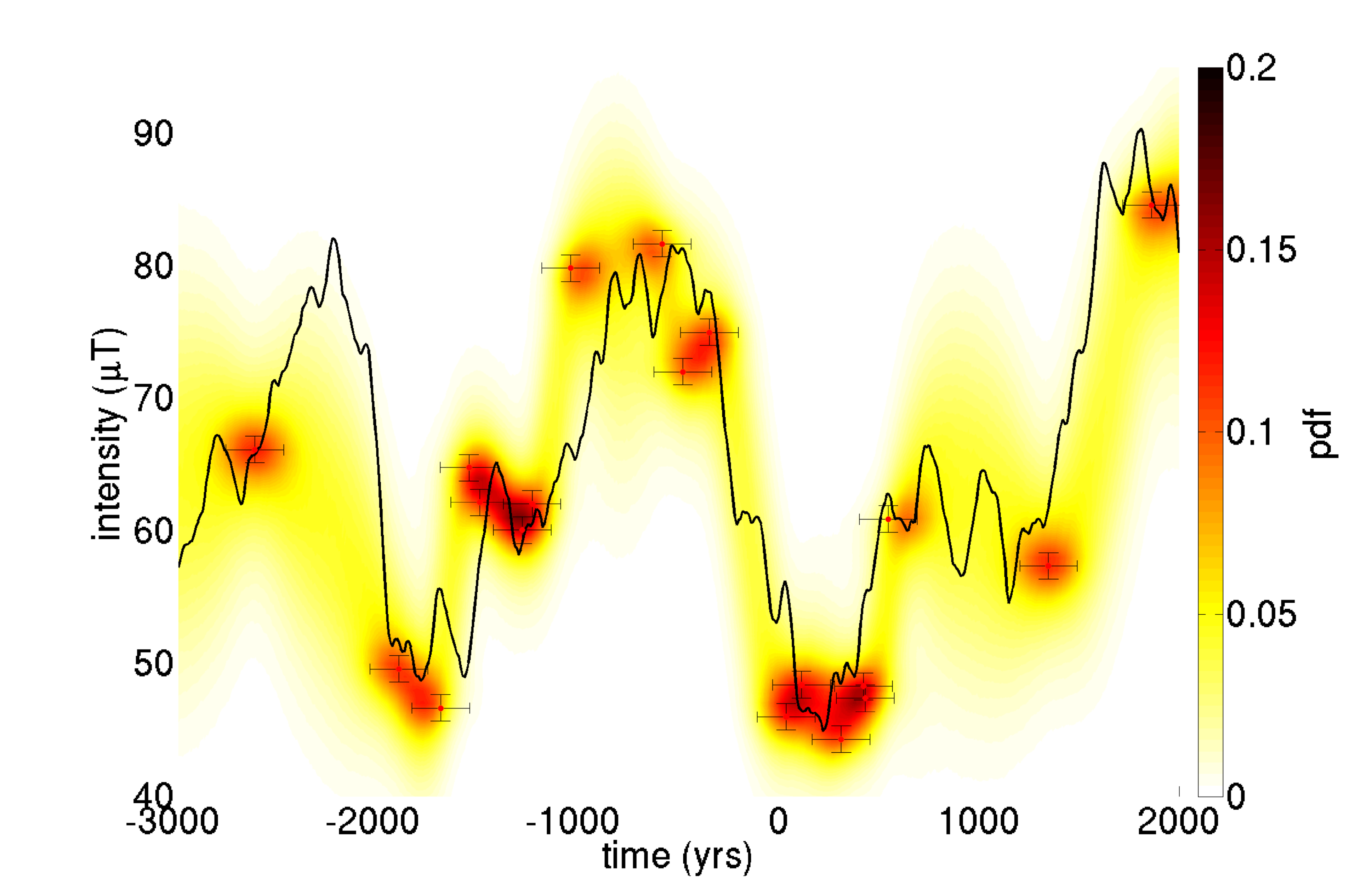}\label{F_gauss}}
\caption{Probability density function of the intensity obtained from a synthetic dataset sampled from a true series (black). a) 20 data, correct errors used in the inversion ; b) 50 data, correct errors used in the inversion ; c) same as (a) but all measurement errors smaller than 5$\mu$T have been converted to 5$\mu$T ; d) same as (a) but with dating errors multiplied by a factor of ten ; e) Same as d) but assuming Gaussian dating errors. The half width of the uniform law $\sigma_u$, has been transformed into the standard deviation of the Gaussian law $\sigma_g=\sigma_u/\sqrt{3}$ so that the uniform and Gaussian law have the same standard deviation. The standard deviation of the Gaussian law and the half width of the uniform law used to generate random measurement and dating errors are set randomly over the dataset with mean values of 1 $\mu$T for measurement errors and 25 yrs (a), (b), and (c) or 250 yrs (d) and (e) for dating uncertainties. \label{F1050}}
\end{center}
\end{figure}

We report in figures \ref{F1050}(a) and \ref{F1050}(b) the obtained \textit{pdf} of the intensity. We first notice that the distribution always encompasses the true series (black curve). 
In the case where 20 data only are available, the sharp changes present in the true series are not closely recovered by the \textit{pdf} due to the lack of data, and the range of estimates is wide except during the few time intervals that are well sampled.
Increasing the quantity of synthetic observations dramatically improves the fit of the \textit{pdf} to the true series and narrows the distribution (see figure \ref{F_50}). 

In figure \ref{F_more}, we invert the same dataset as in figure \ref{F_10}, here again noised following uniform and Gaussian laws with respectively 25 years half-width and 1$\mu$T standard deviation. 
However, following the strategy used by \cite{donadini09}, we assign in the inversion a minimum threshold value for measurement errors, that replaces error estimates lower than this minimal value, chosen to be 5~$\mu T$ for intensities.  
The distribution is  significantly affected by this process, the dispersion happens to be strongly increased particularly when data are available. 
We conclude here that this way of handling small measurement errors penalizes accurate data and leads to lose information.
In figure \ref{F_10times}, we invert the same dataset as in figure \ref{F_10} but after multiplying dating errors by a factor of ten. The dispersion is then a lot wider for the whole studied period. 

For all the precedent cases, the dating errors are supposed uniform what is mostly the case for archeomagnetic objects. However, some of them are dated by radiocarbon methods, which can lead to Gaussian or more complicated error distributions. Figure \ref{F_gauss} presents the obtained \textit{pdf} when dating errors are assumed Gaussian for the inversion of the same dataset as in figure \ref{F_10times}. We see that the resulting \textit{pdf} are rather similar, although Gaussian dating errors slightly increase the \textit{pdf} when observations are available. 
These tests illustrate the importance of assigning realistic error bars for both dating and measurement errors. 
Furthermore, it shows that our method, where the posterior covariance matrix is used to estimate the model error, is capable of accounting for a realistic measure of the information contained into geomagnetic observations, and thus avoids reducing the importance of relatively more accurate records. 

We have evaluated the importance of considering covariances between intensity and inclination within synthetic tests but have not seen significant differences while inverting jointly or separately these observations. Further on, covariances between $F$ and $I$ are considered. 

\newpage
\section{Application to data sets from Syria and France}

In this section, we present results for Mari (Syria) and Paris (France), obtained from intensity data in Syria for epochs between 4000 BC and 0 and from directional and intensity data in France for epochs between 0 and 1900 AD.
For directions, we have converted the 95$\%$ cone of confidence ($\alpha_{95}$) onto declination $\sigma_D$ and inclination $\sigma_I$ errors \citep{piper89}:
%\begin{linenomath*}
\begin{equation}
%\left\{
\begin{array}{rl}
\sigma_D=\displaystyle \frac{81}{140 \cos{I}}\alpha_{95} ; \sigma_I=\displaystyle \frac{81}{140}\alpha_{95}
\end{array}
%\right.\,.
\end{equation}
%\end{linenomath*}

\subsection{Archeointensity data from the Middle-East}

The dataset used here comprises 39 intensity values for Syria \citep{genevey03,gallet06,gallet08,gallet10}. 
All data are reduced to Mari in Syria using the geomagnetic axial dipole (GAD) hypothesis. 
The error caused by the reduction is small compared to measurement errors.
A priori information is built from a magnetic field model truncated  at spherical harmonic degree $N=4$. 
 
\begin{figure}
\begin{center}
\subfigure[Syrian dataset]{\includegraphics[width=0.6\linewidth]{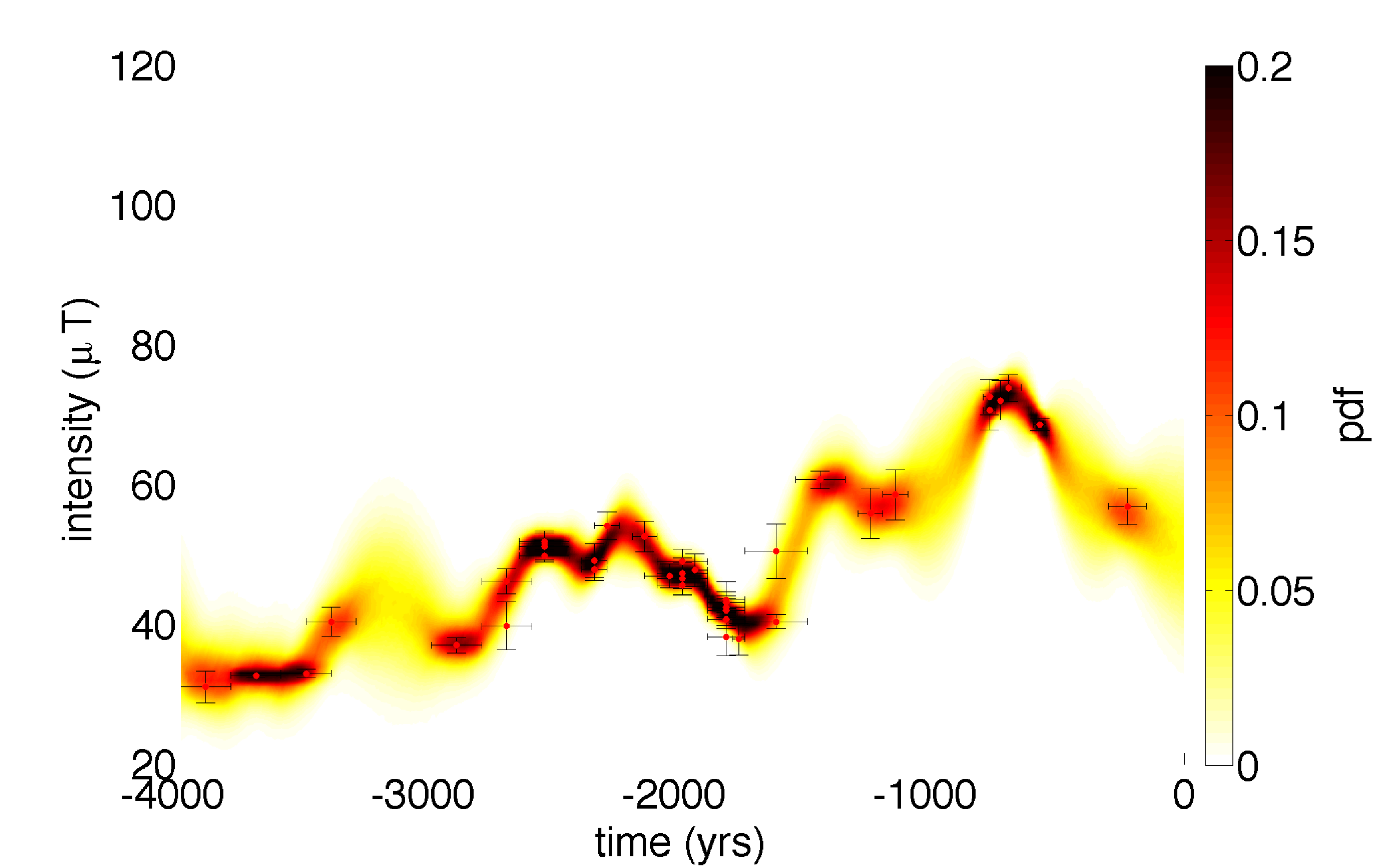}}
\subfigure[Expanded dataset, Huber-norm]{\includegraphics[width=0.6\linewidth]{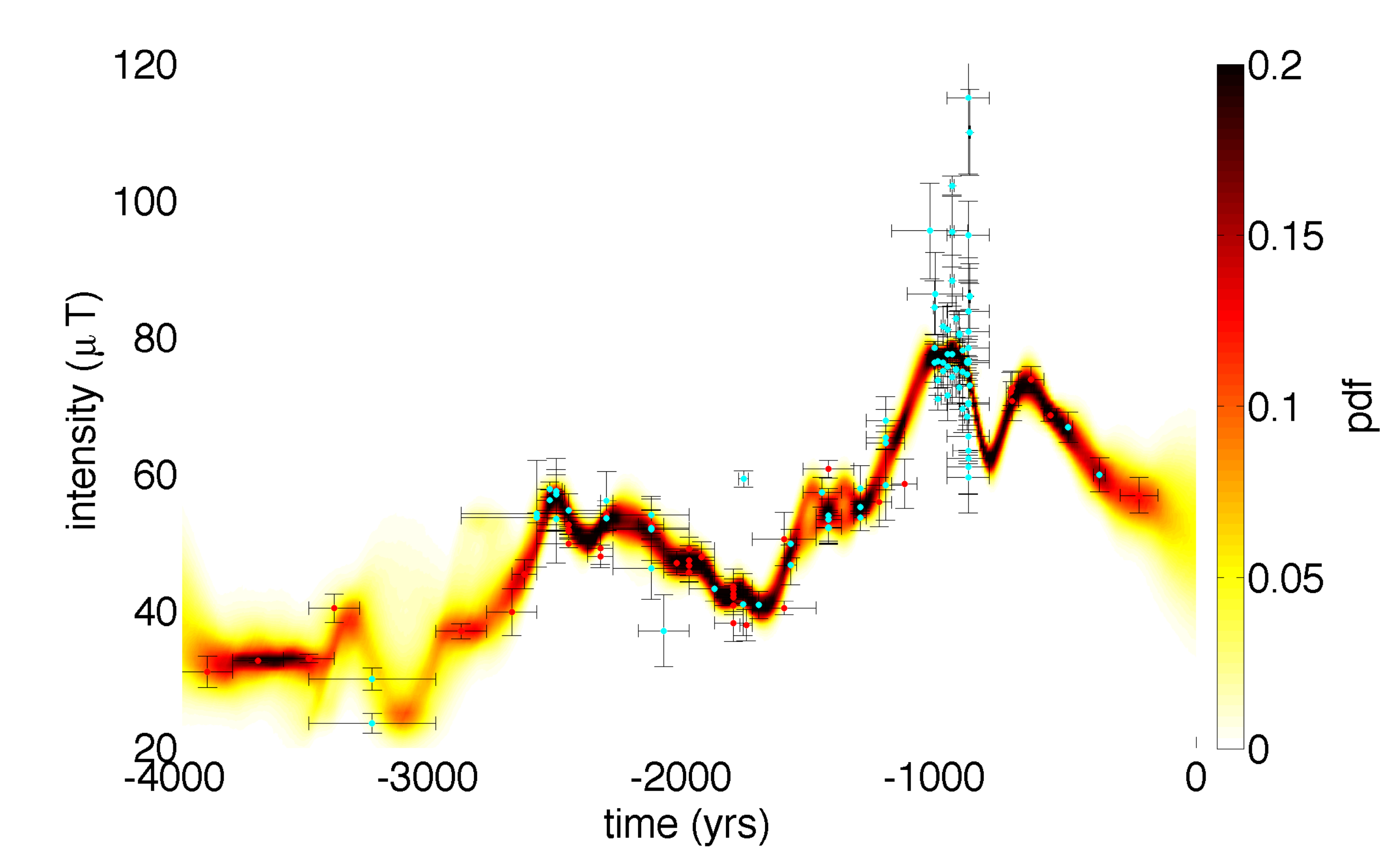}}
\subfigure[Expanded dataset, L2-norm]{\includegraphics[width=0.6\linewidth]{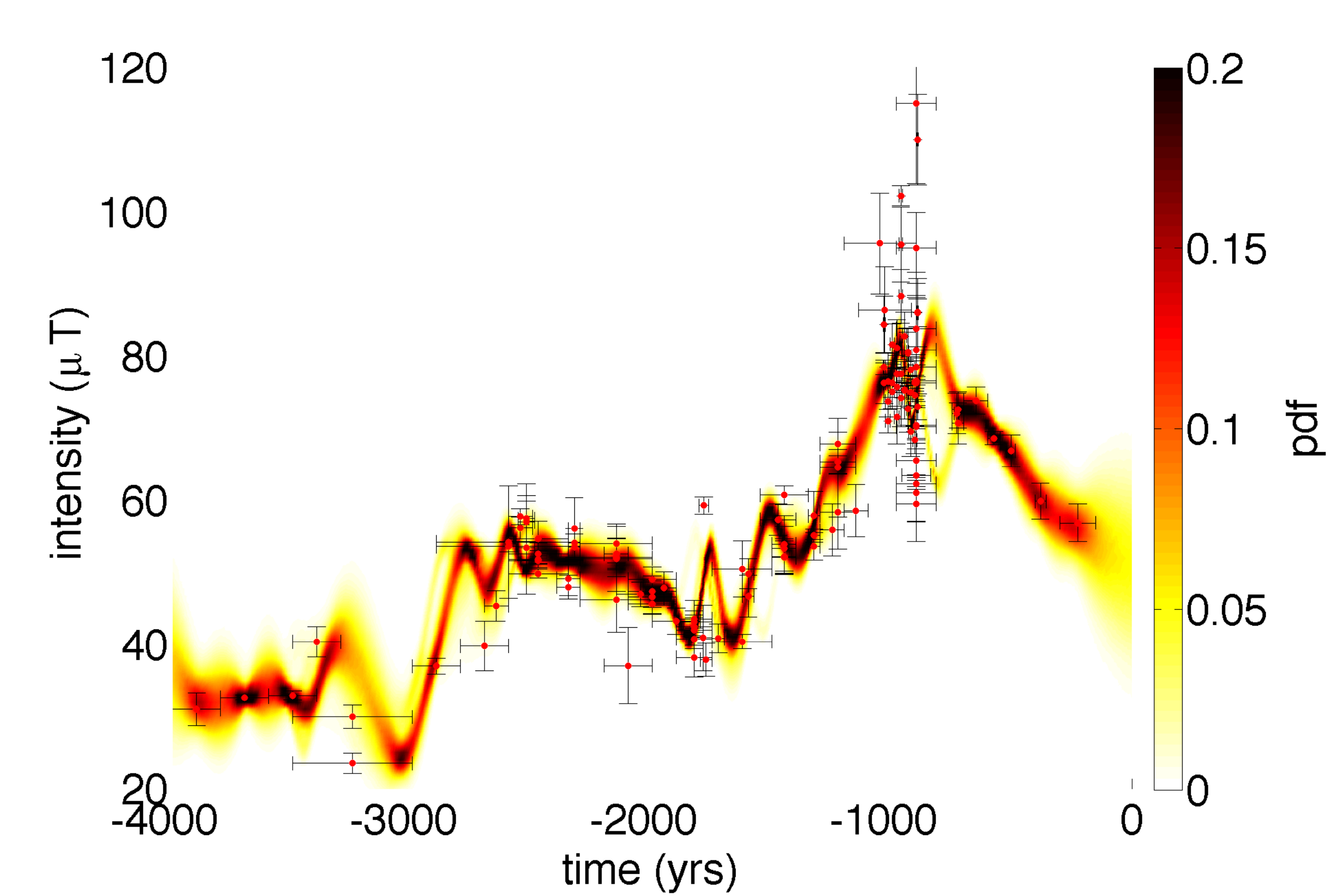}}
\caption{Probability density function from intensity records a) from Syria alone and from the entire Levantine region b) using Huber-norm, c) using L2-norm. Remanence intensities have been transferred to the site of Mari ($34\degr$N, $40\degr$E) via the geomagnetic axial dipole hypothesis. 
\label{syr-ME}}
\end{center}
\end{figure}

Results are displayed in figure \ref{syr-ME}(a) for the Syrian dataset. The distribution is narrow between 2700 and 1600 BC due to the numerous data present during this period. 
Local maxima appear resolved in 2500, 2250, 1450 and 650 BC. 
The distribution prevents us from concluding about extrema value around 3200 BC. Next, we have augmented the dataset with new archeointensity data from Syria \citep{gallet2006high,gallet2014archaeological,gallet2014archaeomagnetism}, and data from the southern Levantine region and Iran \citep{benyosef08,benyosef09,ertepinar12,shaar11}. The new data are plotted in blue. Note that the dataset used here comprises more data than the expanded dataset used by \cite{thebault10}.
Study of the distribution obtained from this expanded dataset confirms the maxima inferred in 2500 and 2250 BC, figure \ref{syr-ME}(b). 
Two sharp maxima appear in 1000 and 650 BC. On figure \ref{syr-ME}a), we remark a wide dispersion around 3200 BC. The few data added between 3500 and 3000 BC, despite very large uncertainties, point to a maximum in 3400 BC followed by a local minimum in 3200 BC although the distribution is still wide. Increasing the number of observations refines the distribution. We see that a mean model in Figure \ref{syr-ME}(a), would not predict the behavior observed with the expanded dataset, the reason why we use probability density functions to represent the results. 

Our modeling strategy differs from the iterative inverse method previously developed by \cite{thebault10}. 
The latter consists in a projection onto cubic B-splines, penalizing the second time derivative, together with a bootstrap strategy to handle dating and measurement errors. Our results for the restricted dataset (figure \ref{syr-ME}(a)) present more rapid variations. Particularly for the two maxima of 2250 and 2500 BC which are well defined in our study and confirmed by the recent observations, whereas the master curve in \cite{thebault10} is flat for this period. In comparison, our distribution presents also a wider dispersion, particularly when data are sparse. 

We show in Figure \ref{syr-ME}(c) the results for the expanded dataset when using the L2-norm.  Sharp local maxima appear now around 800, 1700, 2800 BC which are not apparent in figure \ref{syr-ME}(b). This behavior illustrates a common issue in archeomagnetic modeling. Even when the method accounts for all uncertainties present in the dataset, some incompatibilities within the dataset cannot be handled. One record appearing in 1785 BC has small dating and measurement errors so it forces alone a sharp variation of the model. A rejection criterium has been used in \cite{thebault10} to tackle this issue. We see that the Huber norm alleviates also this difficulty still allowing these data to possibly keep some influence through the MCMC sampling. Here, we show the importance of assigning realistic measurement and dating errors to all data.

\begin{figure}
\begin{center}
\subfigure[3700 BC]{\includegraphics[width=0.32\linewidth]{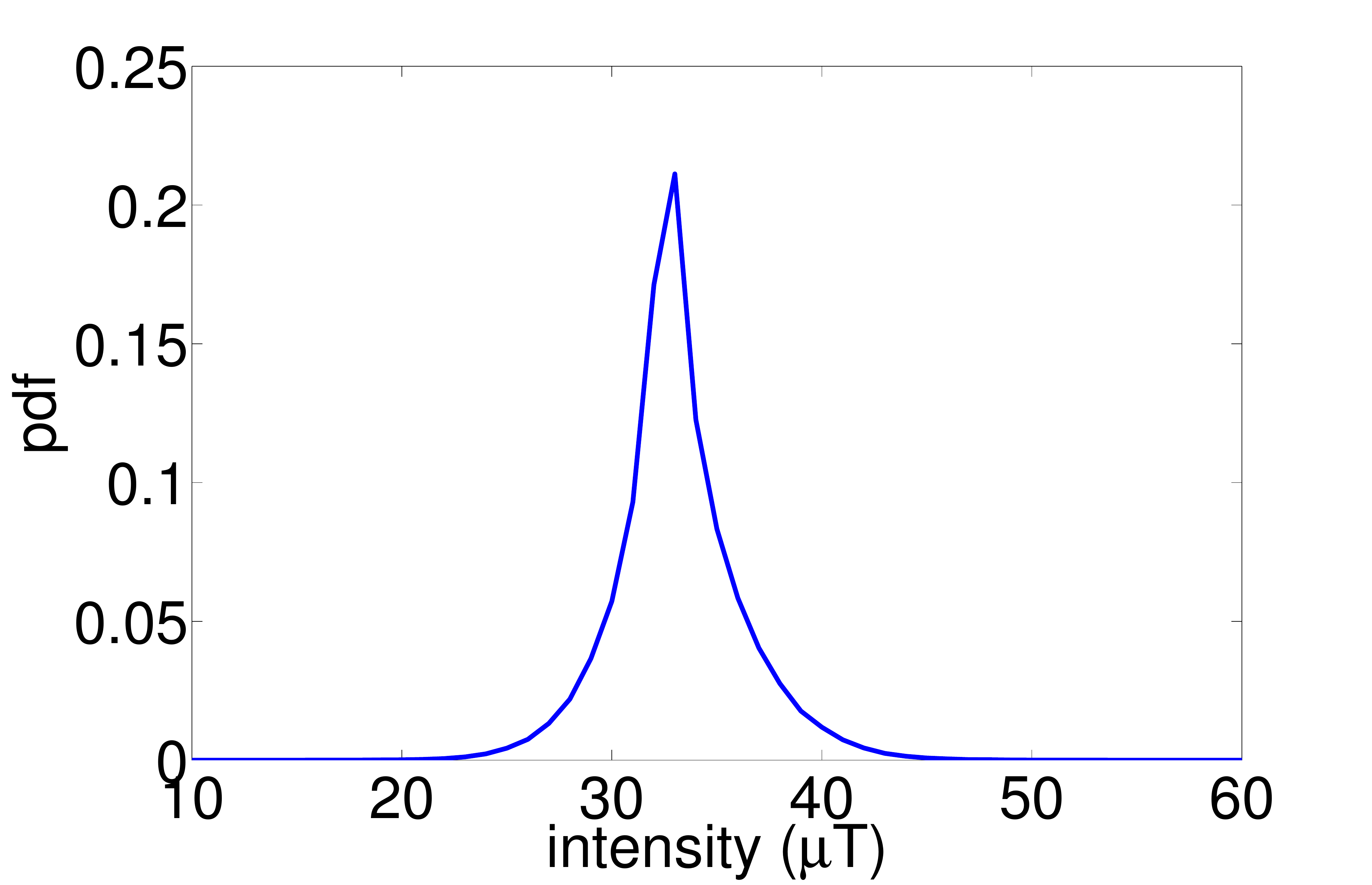}} 
\subfigure[1500 BC]{\includegraphics[width=0.32\linewidth]{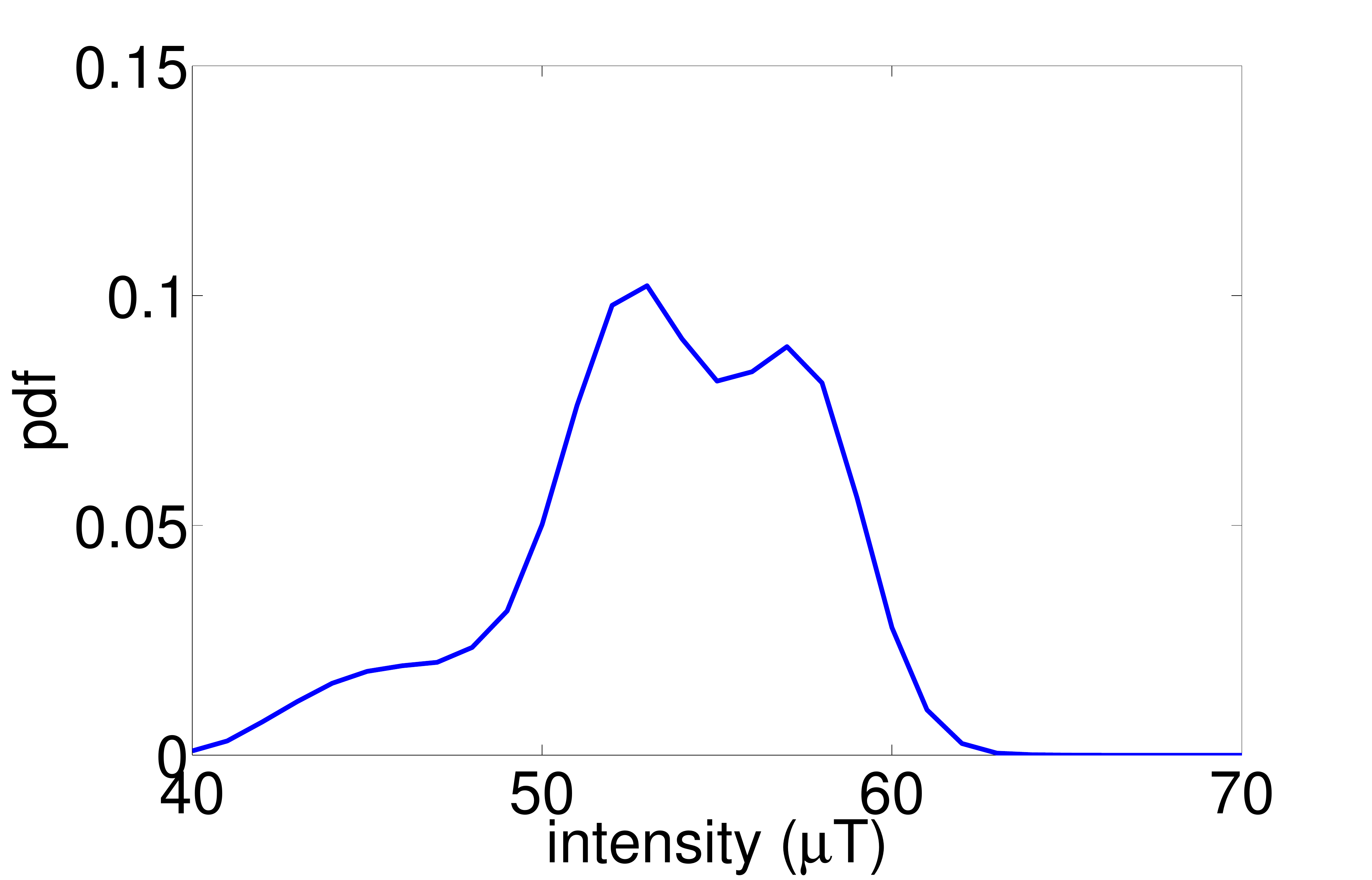}} 
\subfigure[50 BC]{\includegraphics[width=0.32\linewidth]{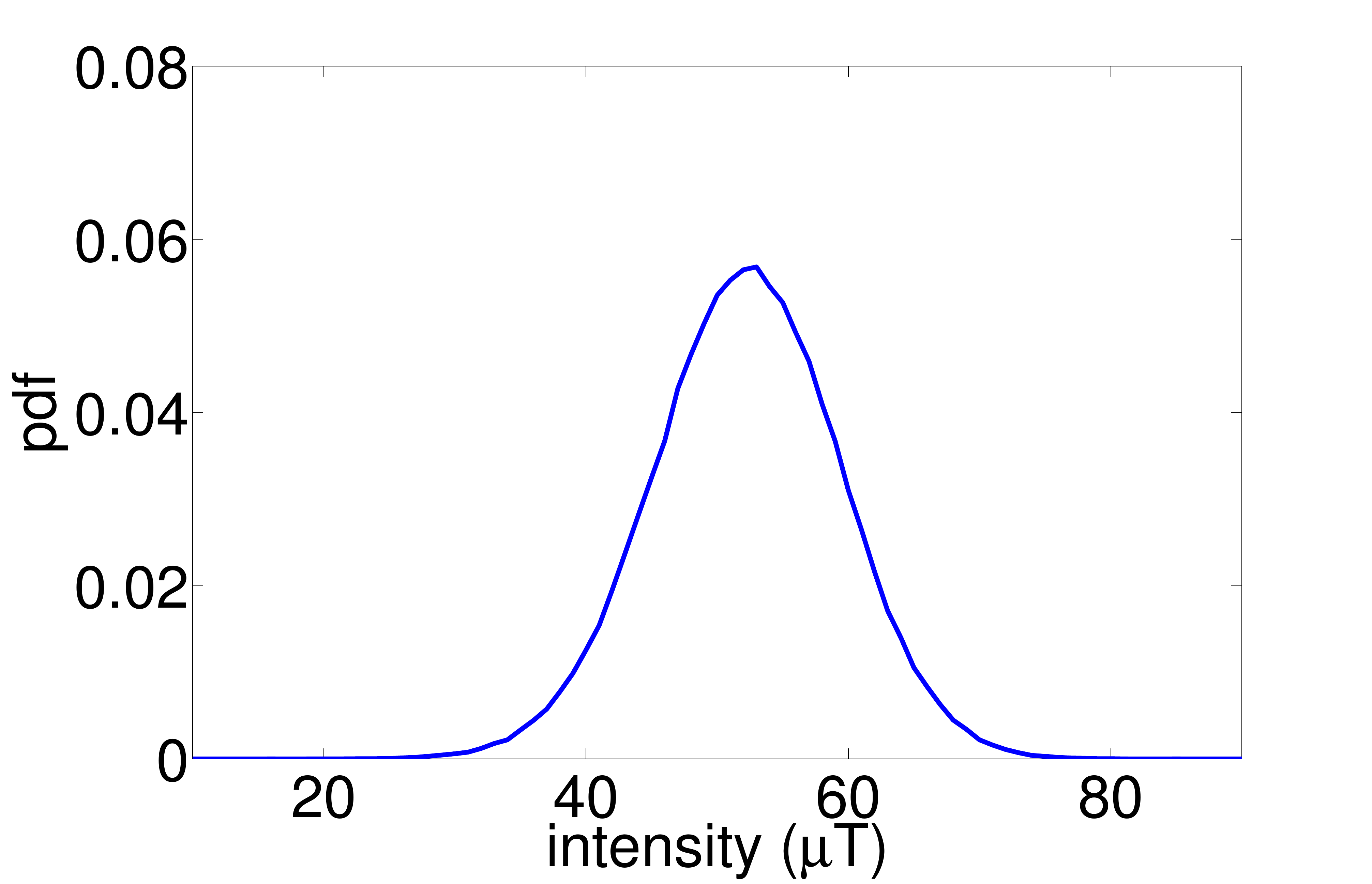}} 
\caption{Probability density functions of the intensity of three cross-sections in 3700, 1500 and 50 BC for the extended dataset.\label{cross}}
\end{center}
\end{figure}

Note also that the posterior distribution is not necessarily Gaussian. Figure \ref{cross} shows three \textit{pdf} of the intensity estimated in 3700, 1500 and 50 BC. We see that the distributions can be similar to Laplacian distribution (\ref{cross}a), Gaussian distribution (\ref{cross}c) or multi-modal distributions (\ref{cross}b). This finding makes awkward the definition of a mean model, the reason why we only consider \textit{pdf} and not master curves.

\begin{figure}
\begin{center}
\subfigure[]{\includegraphics[width=0.32\linewidth]{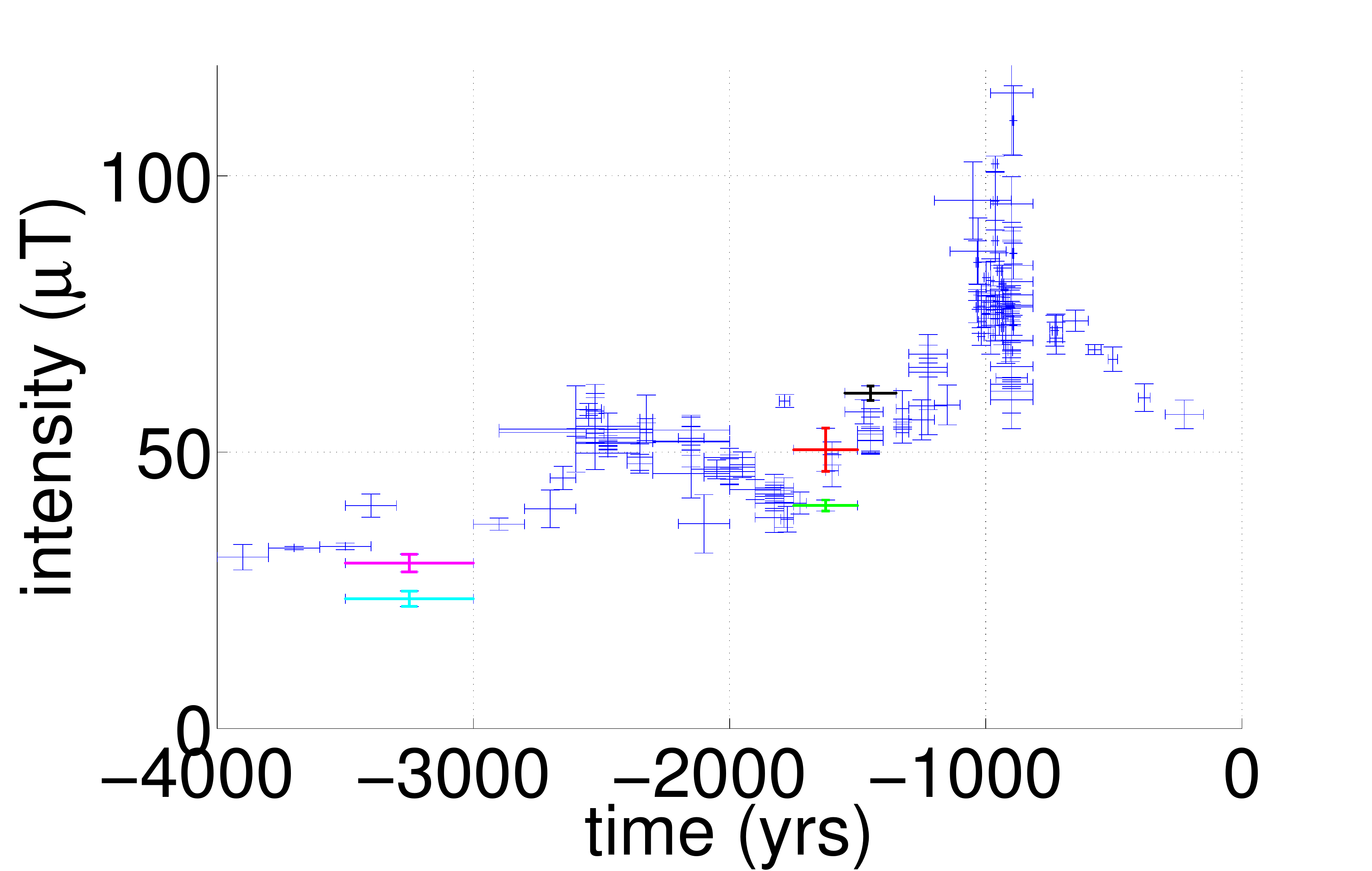}} 
\subfigure[]{\includegraphics[width=0.32\linewidth]{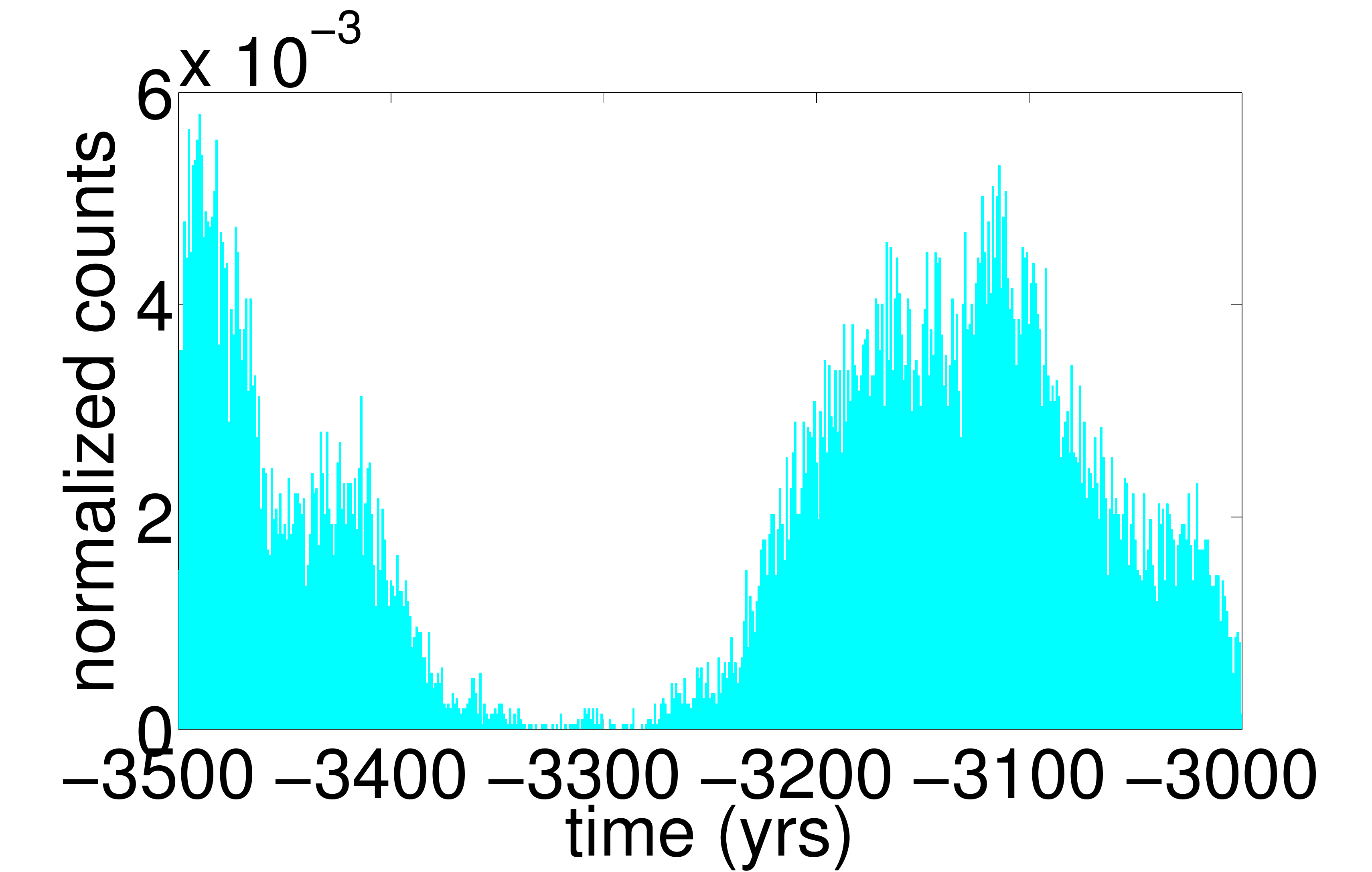}} 
\subfigure[]{\includegraphics[width=0.32\linewidth]{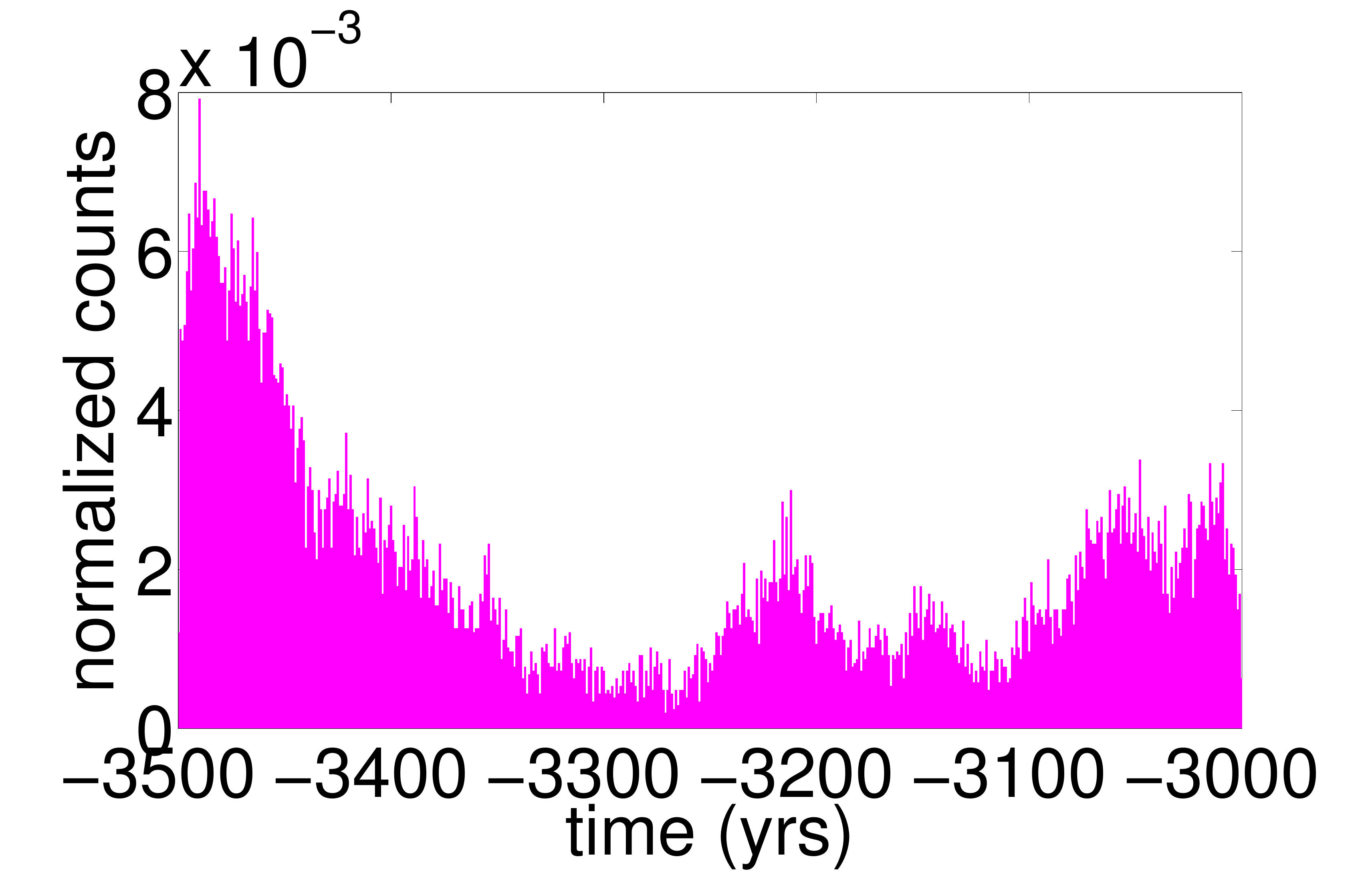}} \\
\subfigure[]{\includegraphics[width=0.32\linewidth]{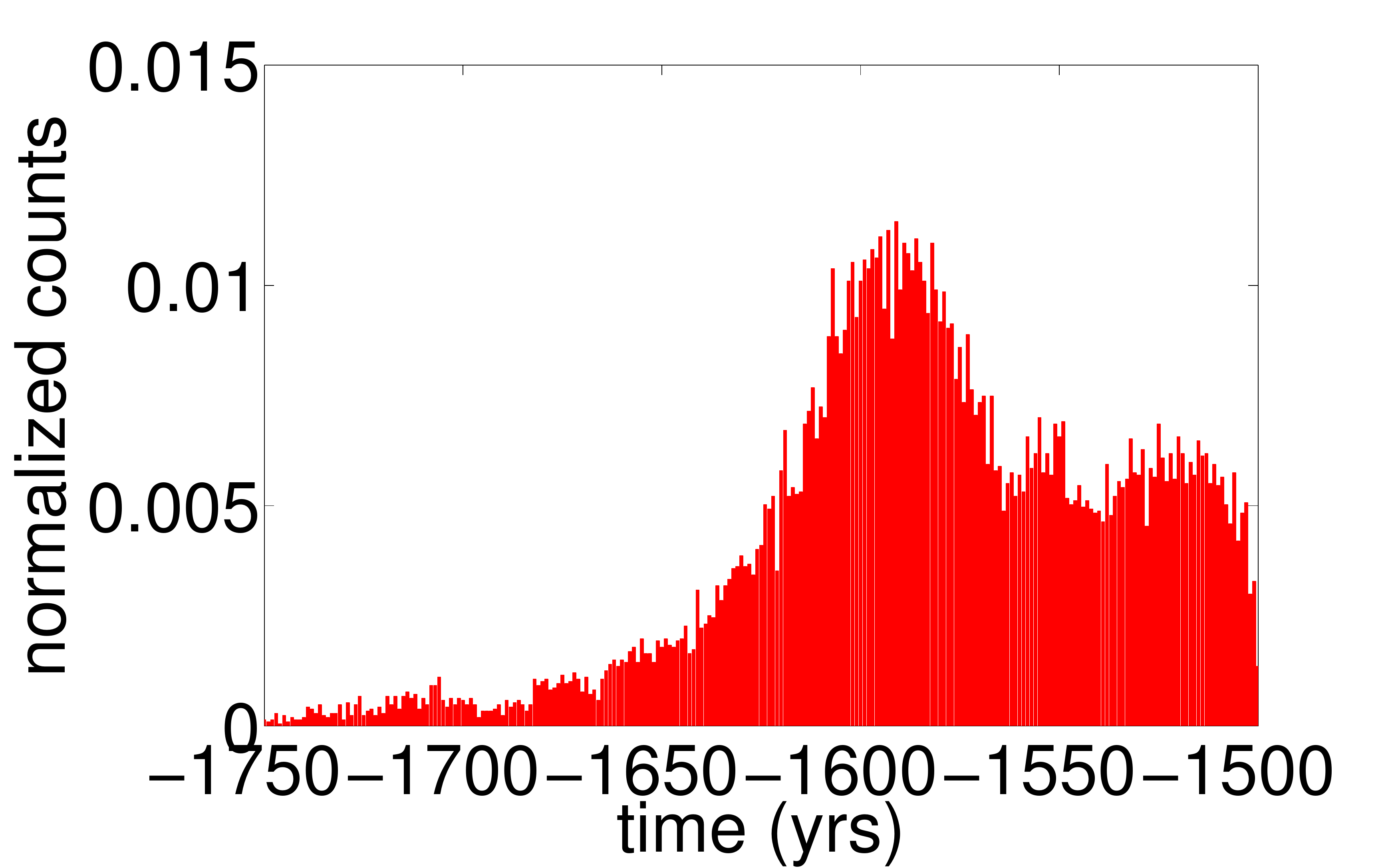}} 
\subfigure[]{\includegraphics[width=0.32\linewidth]{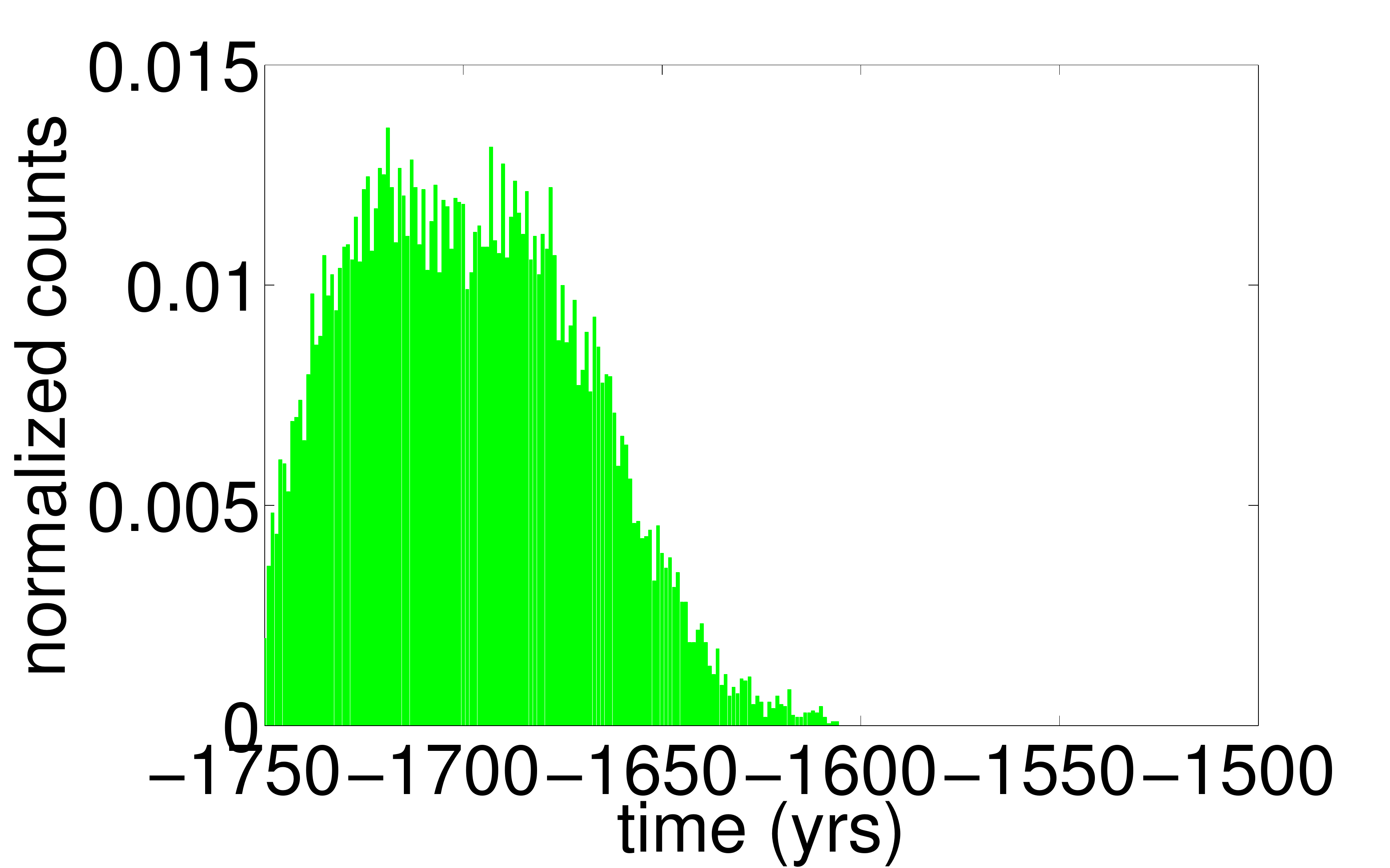}} 
\subfigure[]{\includegraphics[width=0.32\linewidth]{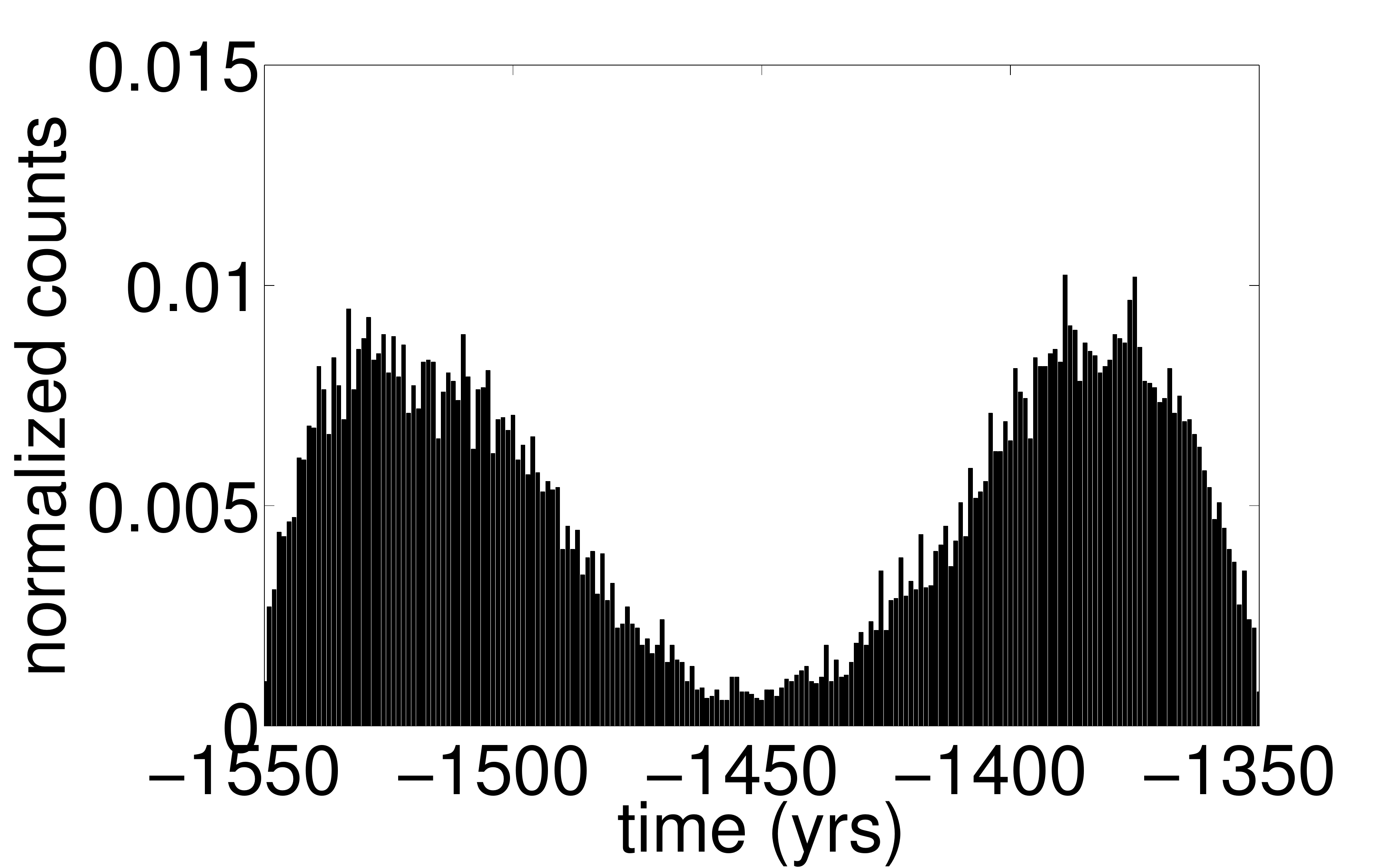}}
\caption{b), c), d), e), f): Distribution of the dates after MCMC colored respectively in a). \label{data}}
\end{center}
\end{figure}

Finally, an important result of our method is the posterior probability on dates. These distributions are very different from their a priori uniform distribution. We focus on five data of the extended dataset (see colored error bars in Figure \ref{data}(a)) and show histograms of the dates preferentially selected in the Markov chains (Figure \ref{data}(b-f)). 
The distribution of dates in figure \ref{data}(d) is very different from a uniform distribution: very few dates appear before 1680 BC and the highest probability for this date is for epochs younger than 1650 BC. Figure \ref{data}(f) displays a multi-modal distribution that makes unlikely epochs around 1450 BC. This methodology can be used to refine the \textit{pdf} of record dates. 
\subsection{Direction and intensity of the magnetic field in Paris}

\begin{figure}
\begin{center}
\subfigure[Declination]{\includegraphics[width=0.6\linewidth]{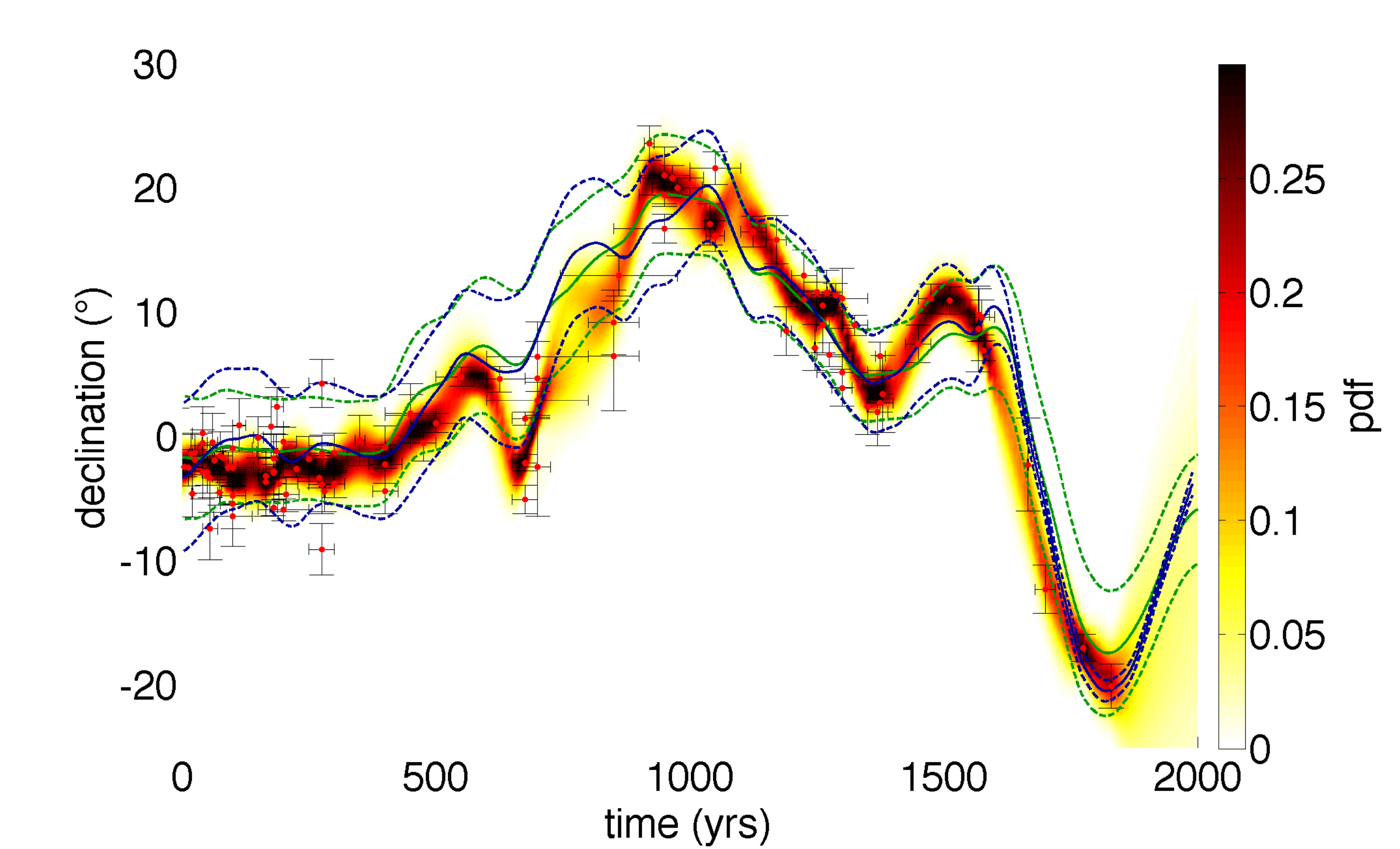}\label{D_paris}}\\
\subfigure[Inclination]{\includegraphics[width=0.6\linewidth]{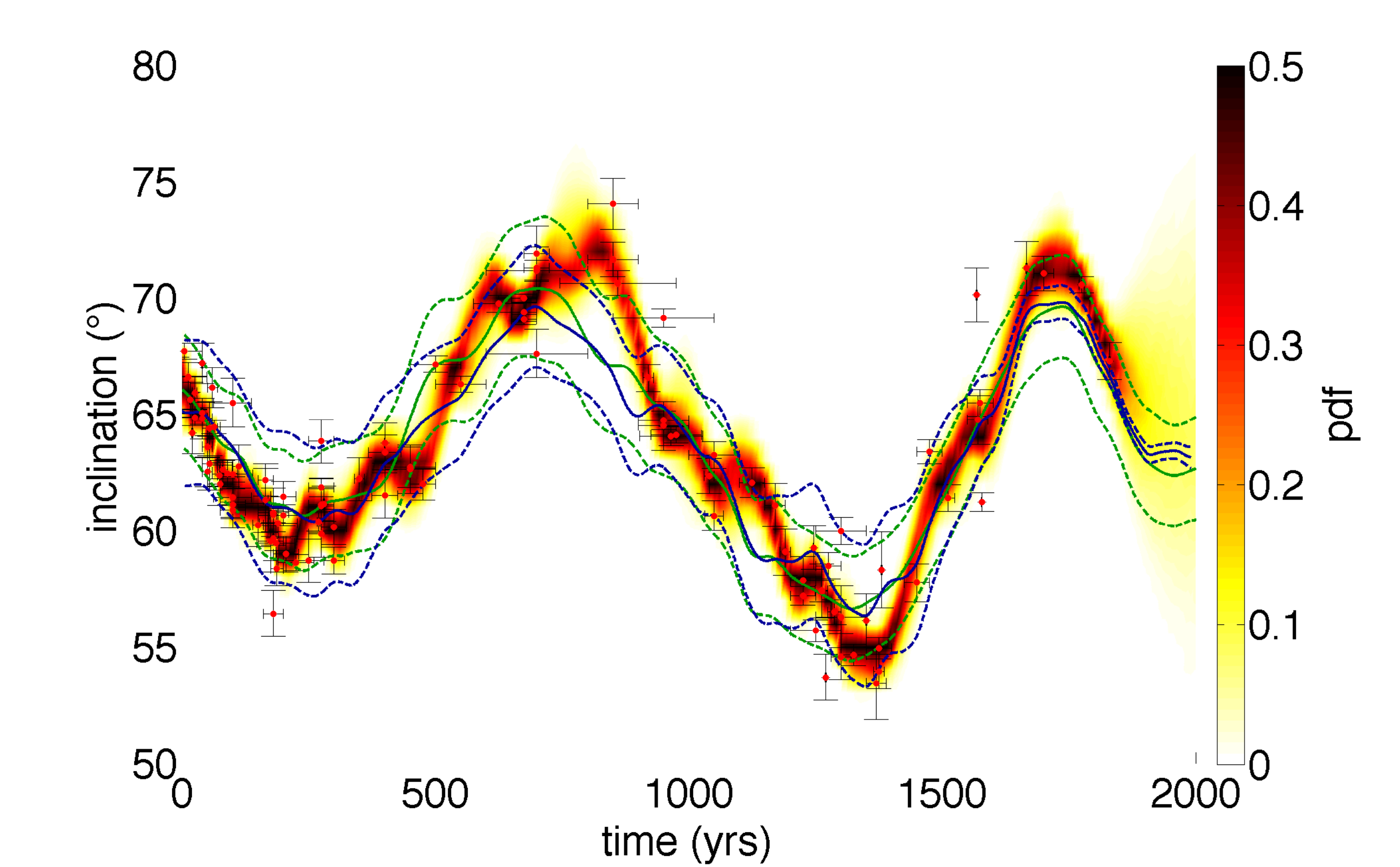}\label{I_paris}} \\
\subfigure[Intensity]{\includegraphics[width=0.6\linewidth]{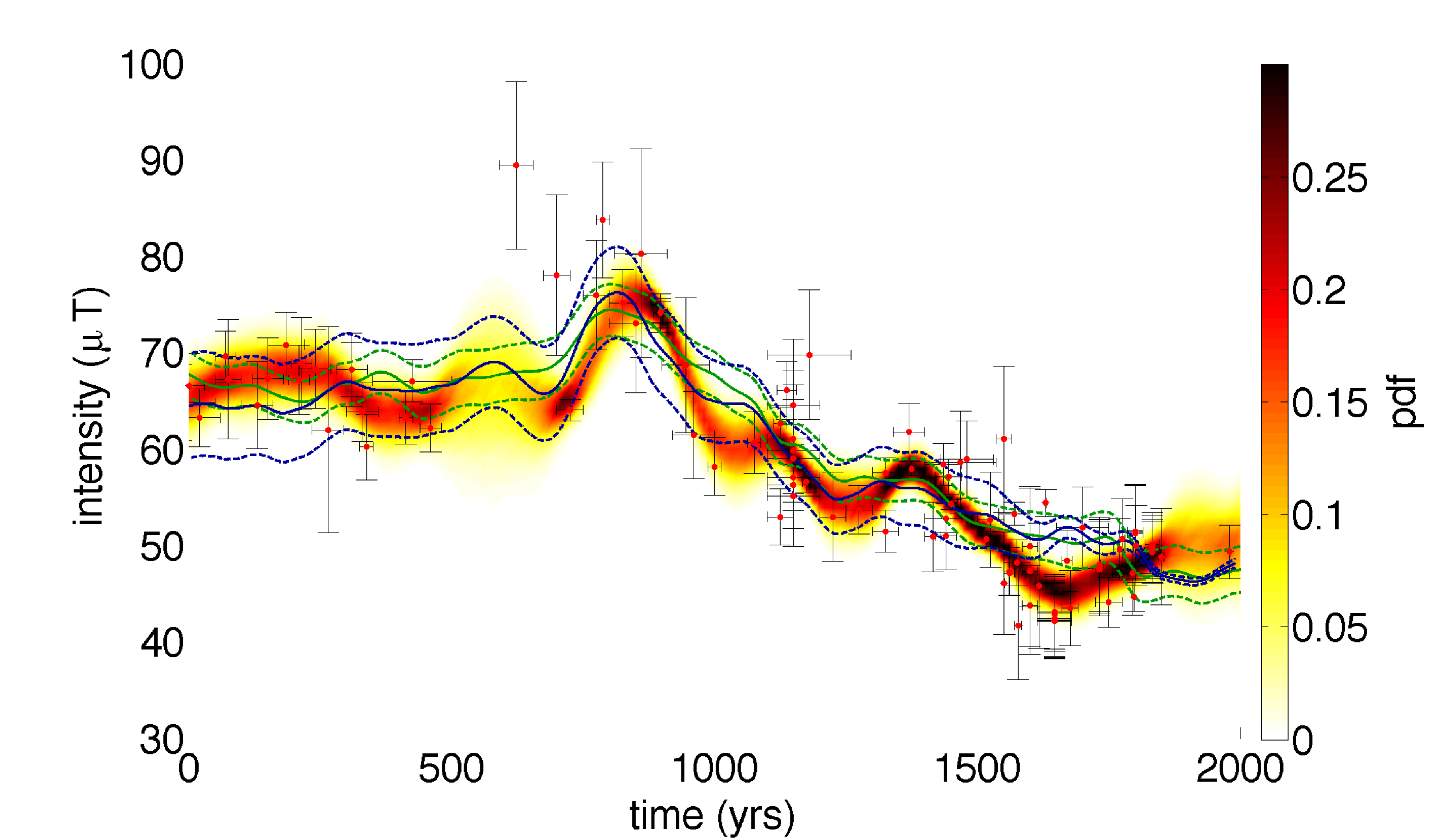}\label{F_paris}}
\caption{Probability density functions of declination, inclination and intensity records from France. All data have been reduced to Paris (48.9$\degr$N, 2.3$\degr$E). The blue curve represents the prediction from ARCH3k and the green curve the prediction from A-FM with their respective 68$\%$ confidence interval (dashed lines).  \label{paris}  }
\end{center}
\end{figure}

We use in this section directional data collated by \cite{bucur94}, and some of the intensity data presented in \cite{genevey13} for France. We adopt the quality criteria of \cite{genevey09}  and keep only data with age uncertainties lower than 100 years, acquired using the Thellier and Thellier method with pTRM-check and with a minimum of three results per site. 
The dataset finally contains 119 directional values and 104 intensity measurements. 
All of them have been reduced to Paris using virtual geomagnetic poles derived from the GAD hypothesis. 
Again, the error caused by the reduction is small compared to measurement errors. MCMC parameters are summarized in Appendix B. We need less chains than for the Syrian study due to the smaller dating errors. 
We display in figure \ref{paris} the \textit{pdf} for $D$, $I$ and $F$.
The intensity series present a general decrease from 850 to 1800 AD, with a local maximum in 1350 AD. 
Data coverage is particularly sparse between 500 and 700 AD, which implies a wide dispersion during this period. 
A maximum close to 80$\mu \text{T}$ appears clearly defined in 850 AD. 
Our results present similar features in comparison with those of \cite{genevey13}, except for the local maximum around 1600 AD that does not exist in our study. 

Predictions from the ARCH3k global model \citep{korte11} and from the A-FM global model \citep{licht13} are superimposed in figure \ref{paris} for comparison, in blue and green respectively. 
The models are in good agreement for declination series except for periods between 600 and 850 AD. 
For inclination however, the high values found at the end of the IX$^{\text{th}}$ century are not accounted for by the ARCH3k and the A-FM models. The intensity minimum found in our study around 1700 AD is not accounted for by the global models.
The intensity maximum appears in both models but is slightly sharper in our model and delayed towards recent epochs. 
This can be due to the penalization of second-time derivatives in the ARCH3k and A-FM models, which may filter out locally well documented rapid variations in order to avoid spurious oscillations elsewhere or to the fact that this model incorporates globally distributed data. 

The a priori information on the model clearly emerges at epoch for which no data are available. In this study, it particularly appears at the end of the studied time interval for declination and inclination. There, the model \textit{pdf} is controlled by the a priori correlation function which ensures the continuity of the first time derivative through the AR-2 process.

\newpage
\section{Conclusion}
\label{sec: conclusion}

In this study, we have developed a new method for the construction of archeomagnetic \textit{pdf} from inclination, declination, and intensity data. 
Our method is based on Gaussian process regression and it incorporates a priori information consistent with the statistics obtained from historical geomagnetic data. Markov Chain Monte Carlo applied on the dates of observations selects random distribution of dates with the highest probabilities. The Huber norm is applied to deal efficiently with outliers. 
This new method has several advantages: 
first it avoids the use of arbitrary regularization, and any unspecified filtering introduced by the projection onto support functions such as cubic B-splines; 
it furthermore allows to account for dating errors in a probabilistic framework. 

We first try our method on synthetic datasets constructed from AR-2 process series. 
Our tests illustrate the importance of using correct estimates of the dating and measurement errors in the inversion in order to optimally recover the a posteriori errors on model parameters. 
They also show that our method is capable of accounting for data displaying disparate accuracies, without losing information contained into the highest quality records. 
The application of this newly developed method to European datasets provides \textit{pdf} that display rapid fluctuations. 
These are less smooth than changes obtained from regularized global \citep[e.g.,][]{korte09} or regional  \citep[e.g.,][]{thebault10} models. 
The \textit{pdf} together with the posterior probability of the record dates may be useful for a purpose of archeomagnetic dating.

We find particularly interesting the use of the MCMC method in order to efficiently explore the space of possible record dates, as we observe that naive random sampling yields largely disparate probabilities for the different sets of dates. We now plan to extend our method to global models. In this context, efficient sampling is crucial. 

In the present study we employ the simplest AR-2 stochastic process that mimics well high frequency variations of the field. Over longer periods, a -2 slope temporal power spectral density has been put forward \citep{panovska2013observed}. Such a slope is consistent with the identification of archeomagnetic jerks \citep{gallet03}. It has motivated the introduction of AR-1 stochastic process in the modeling of long period changes of the magnetic field \citep{brendel07,buffett13}. Alternative AR-2 processes may be employed to represent the two behaviors on short (5-100 years) and long (300-10,000 years) periods. 
Consider for instance the damped oscillator process \citep[eq 2.155']{yaglom04}, governed by stochastic equations depending on two parameters 	and of the general form:
%\begin{linenomath*}
\begin{equation}
d\frac{d\varphi}{dt}  + 2 \alpha
d\varphi  +\omega^2 \varphi dt= d\zeta(t) \,.
\label{damped oscillator}
\end{equation}
%\end{linenomath*}
The Matérn AR-2 process used in this study corresponds to the case $\alpha=\omega$.
Using instead $2\alpha>\omega^2$ one can mimic both the -2 slope temporal power spectral density found for the dipole moment at periods up to approximately $10^5$ yrs from the analysis of geomagnetic records \citep{constable05}, and retrieved in geodynamo simulations \citep{olson12}, and the -4 slope observed at shorter periods. 
This could be an interesting alternative given the cyclic behavior found for the dipole tilt at millennial periods \citep{nilsson11}.

\begin{acknowledgments}

We particularly thank Yves Gallet for providing us with the dataset from the Middle-East and with useful information about archeomagnetic data.
We thank Erwan Thébault and Chris Finlay for useful comments and discussions about modeling and methodology. We finally thank two anonymous reviewers whose detailed comments led to significant improvements of this manuscript. 
This work has been partially supported by the `French Agence Nationale de la Recherche' under the grant ANR-11-BS56-011.

\end{acknowledgments}

\newpage
\bibliographystyle{model2-names}
\bibliography{these}

\newpage
\appendix
\maketitle
\section{$D$, $I$ and $F$ covariances}

\label{app: DIF covar}
In this Appendix, we derive the statistical properties (i.e. mean values and covariances) of the inclination $I$, declination $D$, and intensity $F$ of the magnetic field at a location of longitude $\phi$ and colatitude $\theta$ at the surface of the Earth. We assume that the Gauss coefficients describing the magnetic field are the result of a random stationary process, are characterized by a null mean value (except for the axial dipole, whose mean value is noted $\bar{g}_1^0$), are independent from each other, and have a covariance function that depends only on degree $n$:
\begin{equation}
\mbox{Cov}(g_n^m(t),g_n^m(t+\tau))=\mbox{Cov}(h_n^m(t),h_n^m(t+\tau))=K_n(\tau)
\end{equation}
Such assumptions amount to impose that the statistical properties of the deviation of the magnetic field from an axial dipole are invariant over the surface of the Earth (as demonstrated in \citep{Hulot:2005xy}).\\

We first derive the statistical properties of the north $X$, east $Y$, and downward $Z$ components of the magnetic field. Their expressions (for a truncation degree $N$) at the surface of the Earth are \citep[e.g.][]{Langel:1987ys}:
\begin{equation}
\left\{
\begin{array}{rl}
X(t) = & \displaystyle\sum_{n=1}^N \sum_{m=0}^n \left[g_n^m(t) \cos m\phi + h_n^m(t) \sin m\phi \right] \frac{d P_n^m(\cos \theta)}{d \theta}\\
Y(t)= & \displaystyle\frac{1}{\sin \theta}\sum_{n=1}^N \sum_{m=0}^n m \left[g_n^m(t) \sin m\phi - h_n^m(t) \cos m\phi \right] P_n^m (\cos \theta)\\
Z(t)= & \displaystyle-\sum_{n=1}^N  (n+1) \sum_{m=0}^n \left[g_n^m(t) \cos m\phi + h_n^m(t) \sin m\phi \right] P_n^m (\cos \theta) 
\end{array}
\right.\,.
\end{equation}
Because only $g_1^0$ has a non-zero mean value, the mean values of $X$, $Y$, and $Z$ are simply
\begin{equation}
\overline{X}=-\overline{g}_1^0\sin\theta\;; \quad \overline{Y}=0\;; \quad \overline{Z}=-2\overline{g}_1^0\cos\theta\,.
\label{meancomponent}
\end{equation}

Because of the independence of Gauss coefficients, and because their covariance function depends only on the spherical harmonic degree $n$, covariances on $X$, $Y$ and $Z$ simplify into :
\begin{equation}
\left\{
\begin{array}{rl}
\mbox{Cov}(X(t),X(t+\tau)) & = \displaystyle\sum_{n=1}^NK_n(\tau)\sum_{m=0}^n\left(\frac{dP_n^m(\cos\theta)}{d\theta}\right)^2\\
\mbox{Cov}(Y(t),Y(t+\tau)) & = \displaystyle\frac{1}{\sin^2\theta}\sum_{n=1}^NK_n(\tau)\sum_{m=0}^nm^2\left(P_n^m(\cos\theta)\right)^2\\
\mbox{Cov}(Z(t),Z(t+\tau)) & = \displaystyle\sum_{n=1}^N(n+1)^2K_n(\tau)\sum_{m=0}^n\left(P_n^m(\cos\theta)\right)^2\\
\mbox{Cov}(X(t),Y(t+\tau)) & = 0 \\
\mbox{Cov}(Y(t),Z(t+\tau)) & = 0 \\
\mbox{Cov}(X(t),Z(t+\tau)) & = \displaystyle- \sum_{n=1}^N(n+1)K_n(\tau)\sum_{m=0}^n\frac{dP_n^m(\cos\theta)}{d\theta}P_n^m(\cos\theta)
\end{array}
\right.\,.
\end{equation}
Such expressions can be further simplified using the following relations for Schmidt normalized associated Legendre functions \citep{winch1995derivatives}: 
\begin{equation}
\left\{
\begin{array}{l}
\displaystyle\sum_{m=0}^n \left(P_n^m(\cos\theta)\right)^2=1\\
\displaystyle\sum_{m=0}^n \left(\frac{dP_n^m(\cos\theta)}{d\theta}\right)^2=\frac{n(n+1)}{2}\\
\displaystyle\sum_{m=0}^n \left(\frac{m}{\sin\theta}P_n^m(\cos\theta)\right)^2=\frac{n(n+1)}{2}\\
\displaystyle\sum_{m=0}^n \frac{dP_n^m(\cos\theta)}{d\theta}P_n^m(\cos\theta)=0
\end{array}
\right.\,.
\end{equation}
We therefore deduce that :
\begin{equation}
\left\{
\begin{array}{rl}
\mbox{Cov}(X(t),X(t+\tau)) & = \displaystyle\sum_{n=1}^N\frac{n(n+1)}{2}K_n(\tau)\\
\mbox{Cov}(Y(t),Y(t+\tau)) & = \displaystyle\sum_{n=1}^N\frac{n(n+1)}{2}K_n(\tau)\\
\mbox{Cov}(Z(t),Z(t+\tau)) & = \displaystyle\sum_{n=1}^N(n+1)^2K_n(\tau)
\end{array}
\right.\,,
\label{cov XYZ}
\end{equation}
and that the series of $X$, $Y$, and $Z$ recorded at a same location are independent from each other :
\begin{equation}
\mbox{Cov}(X(t),Y(t+\tau)) = \mbox{Cov}(Y(t),Z(t+\tau)) = \mbox{Cov}(X(t),Z(t+\tau)) = 0
\end{equation}

The declination $D$, inclination $I$ and intensity $F$ of the magnetic field are not linearly related to the components $X$, $Y$, and $Z$: 
\begin{equation}
D=\arctan{\frac{Y}{X}}\;; \quad I=\arctan\frac{Z}{\sqrt{X^2+Y^2}}\;; \quad F=\sqrt{X^2+Y^2+Z^2}\,.
\label{dif}
\end{equation}
Let us denote $\mathbf{A}=(X,Y,Z)$ and $\mathbf{B}=(D,I,F)$. If the vector $\mathbf{A}$ does not depart much from its mean value $\overline{\mathbf{A}}$ (corresponding to the mean axial dipole), the above non-linear relations, noted $\mathbf{B}=\psi(\mathbf{A})$, can be approximated using a first-order Taylor expansion :
\begin{equation}
B_k=\psi_k(\overline{\mathbf{A}})+\sum_i \left. \frac{\partial{\psi_k}}{\partial{A_i}}\right|_{\overline{\mathbf{A}}}(A_i-\overline{A})\,.
\end{equation}
The mean value of $\mathbf{B}$ is therefore approximated by:
\begin{equation}
\overline{B}_{k}=\psi_k(\overline{\mathbf{A}})
\label{meanb}
\end{equation}
Combining equations (\ref{meanb}), (\ref{meancomponent}), and (\ref{dif}), we obtain the expression for the mean value of $D$, $I$, and $F$:
\begin{equation}
\overline{D}= 0 \text{ (} \pi \text{ if } \overline{g}_1^0 > 0)\;;
\quad \overline{I}=-\mbox{sgn}(\overline{g}_1^0)\arctan\left(\frac{2}{\tan\theta}\right) \;;
\quad \overline{F}=|\overline{g}_1^0|\sqrt{1+3\cos^2\theta}\,,
\end{equation}
and the covariance matrix for $\mathbf{B}$ is approximated by:
\begin{equation}
\mbox{Cov}(B_k(t),B_l(t+\tau))=\sum_i \sum_j \left. \frac{\partial\psi_k}{\partial A_i}\right|_{\overline{\mathbf{A}}}\left. \frac{\partial\psi_l}{\partial A_j}\right|_{\overline{\mathbf{A}}} \mbox{Cov}(A_i(t),A_j(t+\tau))\,.
\end{equation}
Because the series of $X$, $Y$, and $Z$ are independent of each other, this expression can be simplified into:
\begin{equation}
\mbox{Cov}(B_k(t),B_l(t+\tau))=\sum_i \left( \left. \frac{\partial\psi_k}{\partial A_i}\right|_{\overline{\mathbf{A}}}\right)^2 \mbox{Cov}(A_i(t),A_i(t+\tau))\,.
\label{covb}
\end{equation}
This expression involves the partial derivative of $D$, $I$, and $F$ with respect to $X$, $Y$, and $Z$ evaluated at $(\overline{X},\overline{Y},\overline{Z})$ :
\begin{equation}
\left\{
\begin{array}{lll}
\displaystyle\left. \frac{\partial D}{\partial X} \right|_{\overline{{\bf A}}}  = 0  & ; \;
\displaystyle\left. \frac{\partial D}{\partial Y} \right|_{\overline{{\bf A}}}  = -\frac{1}{\overline{g}_1^0\displaystyle\sin\theta} &;   \;
\displaystyle\left. \frac{\partial D}{\partial Z} \right|_{\overline{{\bf A}}}  = 0 \\
\displaystyle\left. \frac{\partial I}{\partial X} \right|_{\overline{{\bf A}}}  = - \frac{2\cos\theta}{|\overline{g}_1^0|(1+3\cos^2\theta)} &;  \;
\displaystyle\left. \frac{\partial I}{\partial Y} \right|_{\overline{{\bf A}}}  = 0 &;  \;
\displaystyle\left. \frac{\partial I}{\partial Z} \right|_{\overline{{\bf A}}}  = \frac{\sin\theta}{|\overline{g}_1^0|(1+3\cos^2\theta)}\\
\displaystyle\left. \frac{\partial F}{\partial X} \right|_{\overline{{\bf A}}}  = -\frac{\mbox{sgn}(\overline{g}_1^0)\sin\theta}{\sqrt{1+3\cos^2\theta}} & ; \;
\displaystyle\left. \frac{\partial F}{\partial Y} \right|_{\overline{{\bf A}}}  = 0 &;  \;
\displaystyle\left. \frac{\partial F}{\partial Z} \right|_{\overline{{\bf A}}}  = -\frac{2\mbox{sgn}(\overline{g}_1^0)\cos\theta}{\sqrt{1+3\cos^2\theta}}
\end{array}
\right.\,.
\label{partial DIF}
\end{equation}
Finally, combining equations (\ref{covb}), (\ref{cov XYZ}), and (\ref{partial DIF}), we obtain the following approximated expressions for the covariances of $D$, $I$, and $F$ :
\begin{equation}
\left\{
\begin{array}{ll}
\mbox{Cov}(D(t),D(t+\tau)) &= \displaystyle\frac{1}{2(\overline{g}_1^0)^2\sin^2\theta} \sum_{n=1}^N {n(n+1)} K_n(\tau) \\
\mbox{Cov}(I(t),I(t+\tau)) &= \displaystyle\frac{2\cos^2\theta}{(\overline{g}_1^0)^2(1+3\cos^2\theta)^2} \sum_{n=1}^N{n(n+1)}K_n(\tau) + \frac{\sin^2\theta}{(\overline{g}_1^0)^2(1+3\cos^2\theta)^2}\sum_{n=1}^N (n+1)^2K_n(\tau) \\
\mbox{Cov}(F(t),F(t+\tau)) &=\displaystyle \frac{\sin^2\theta}{2(1+3 \cos^2\theta)} \sum_{n=1}^N{n(n+1)}K_n(\tau)+ \frac{4 \cos^2\theta}{1+3 \cos^2\theta} \sum_{n=1}^N (n+1)^2K_n(\tau)
\end{array}
\right.\,,
\label{covar DIF app}
\end{equation}
and the cross-covariances within the different quantities:
\begin{equation}
\left\{
\begin{array}{rl}
\mbox{Cov}(D(t),I(t+\tau)) =& 0\\
\mbox{Cov}(I(t),F(t+\tau)) =& 0\\
\mbox{Cov}(I(t),F(t+\tau)) =&\displaystyle -\frac{ \cos\theta \sin\theta}{\overline{g}_1^0 (1+3 \cos^2\theta)^{3/2}}\sum_{n=1}^N {(n+1)(n+2)} K_n(\tau)
\end{array}
\right.\,.
\label{cross-covar DIF app}
\end{equation}

\newpage

\maketitle
\section{Parameters used in the MCMC method. Number $N$ of iterations per chain, $\sigma_{\text{MCMC}}$ and number  $N_{\text{MCMC}}$ of draws selected by the Markov rules. The number of lines corresponds to the number of chains used for each figure.}
\begin{center}
\begin{tabular}{c c c c || c c c c}
  & $N$ &$\sigma_{\text{MCMC}}$ & $N_\text{MCMC}$ & & $N$ & $\sigma_{\text{MCMC}}$ &$N_\text{MCMC}$\\
 Fig. 3(a) & 10000 & 50&6926&&10000&5&3348\\
  &10000&50&6953&&10000&5&3424\\
  Fig. 3(b)&10000&10&3821 & Fig. 4(c)&10000&1&3200\\
   &10000&10&3820&&10000&1&2830\\
   &10000&10&3777&&10000&1&2050\\
   Fig. 3(c)&1000&100&962&&10000&1&3157\\
   &1000&100&970&&10000&1&2825\\
   &1000&100&962&&10000&1&3856\\
   Fig. 3(d)&10000&50&4439&&10000&1&3812\\
   &10000&50&4442&&10000&1&3132\\
   &10000&50&4437&&10000&1&3156\\
   &10000&50&4442&&10000&1&3166\\
   Fig. 3(e)&10000&50&4351&Fig. 7(a)&10000&2&5000\\
   &10000&50&4392&&10000&2&5170\\
   &10000&50&4268&&10000&2&5002\\
   &10000&50&4463&&10000&2&4718\\      
   Fig. 4(a)&10000&30&2980&Figs. 7(b and c)&10000&2&3990\\
   &10000&30&2950& &10000&2&3180\\
   &10000&30&3050& &10000&2&2480\\
   &10000&30&3090&  &10000&2&3480\\
   &10000&30&3010& &10000&2&3492\\
   Fig. 4(b)&10000&5&3379\\
   &10000&5&3491\\
   &10000&5&3576\\
   &10000&5&3062\\
   &10000&5&3622\\
   &10000&5&3593\\
   &10000&5&3516\\
   &10000&5&3330\\
   &10000&5&3132\\
   &10000&5&3166\\
   &10000&5&3348\\
    &10000&5&3424\\

   \hline
\end{tabular}

\end{center}

\end{document}